\documentclass[aps, prd, amsmath, amssymb, floats, floatfix, superscriptaddress, nofootinbib, twocolumn, showpacs,reprint]{revtex4-1}

\usepackage{graphicx}
\usepackage{amsmath}
\usepackage{amssymb}
\usepackage{amsfonts}
\usepackage{times}
\usepackage{xspace} 
\usepackage[usenames]{color}
\usepackage{dcolumn}
\usepackage{bm}
\usepackage{mathrsfs}
\usepackage{tikz}
\usepackage{appendix}

\usepackage{ulem}
\usepackage{verbatim}
\usepackage{appendix}
\usepackage{lipsum}

\begin{document}

\title{Post-Newtonian constraints on Lorentz-violating gravity theories with a MOND phenomenology}

\author{Matteo Bonetti}
\affiliation{Dipartimento di Fisica G. Occhialini, Universit\`a degli Studi di Milano Bicocca,\\ Piazza della Scienza 3, 20126 Milano, Italy}
\affiliation{Dipartimento di Scienza e alta Tecnologia, Universit\`a degli Studi dell'Insubria,\\ Via Valleggio 11, 22100 Como, Italy}
\author{Enrico Barausse}
\affiliation{CNRS, UMR 7095, Institut d'Astrophysique de Paris, 98bis Bd Arago, 75014 Paris, France}
\affiliation{Sorbonne Universit\'es, UPMC Univ Paris 06, UMR 7095, 98bis Bd Arago, 75014 Paris, France}

\begin{abstract}
We study the post-Newtonian expansion of a class of Lorentz-violating gravity theories that reduce to khronometric theory (i.e. the infrared limit of Ho\v rava gravity)
in high-acceleration regimes, and reproduce the phenomenology of the modified Newtonian dynamics (MOND) in the low-acceleration, non-relativistic limit. 
Like in khronometric theory, Lorentz symmetry is violated in these theories by introducing a dynamical scalar field (the ``khronon'') whose gradient is enforced to be timelike. As a result, hypersurfaces of constant khronon define a preferred foliation of the spacetime, and the khronon can be thought of as a physical absolute time.
The MOND phenomenology arises as a result of the presence, in the action, of terms depending on the acceleration of the congruence orthogonal to the preferred foliation. 
We find that if the theory is forced to reduce exactly to General
Relativity (rather than to khronometric theory) in the
high-acceleration regime, the post-Newtonian expansion breaks down at
low accelerations, and the theory becomes strongly coupled.
Nevertheless, we identify a sizeable region of the parameter space where the post-Newtonian expansion remains perturbative for all accelerations, and the theory passes both solar-system and pulsar gravity tests, besides producing a MOND phenomenology for the rotation curves of galaxies. We illustrate this explicitly 
with  a toy model of a system containing only baryonic matter but no Dark Matter.
\end{abstract}

\date{\today}

\maketitle

\section{Introduction}
\label{sec:introduction}

The year 2015 marks the hundredth anniversary of General Relativity (GR). This elegant theory has been greatly successful at interpreting and predicting gravitational
phenomena on a huge range of length-scales, velocities, gravitational-field strengths and space-time curvatures. On small length-scales, submillimeter experiments verified the validity of Newtonian gravity, to which GR reduces in the quasi-static weak-field regime characterizing these experiments, down to micro-meter scales~\cite{Adelberger:2006dh,Kapner:2006si}. Newtonian gravity has been historically tested in the solar system, but in the course of the twentieth century technological progress made it possible to test also the first post-Newtonian (1PN) corrections to Newtonian dynamics~\cite{Will:2014xja,Will:1993ns}, i.e. the GR corrections of fractional order ${\cal O}(v/c)^2$, with $v$ being the system's characteristic velocity. Indeed, these 1PN solar-system tests date back to the first triumph of GR, i.e. Einstein's prediction of the correct perihelion shift for Mercury, and later came to include also light-deflection measurements, time-delay and gyroscopical-precession experiments, as well as exquisite verifications of one of GR's building blocks, i.e.~the equivalence principle. However, because velocities in the solar system are $v\lesssim 10^{-4} c$, and the gravitational fields are weak (i.e. $\phi_N/c^2 \lesssim10^{-6}$, $\phi_N$ being the Newtonian potential), tests of the GR dynamics beyond this weak-field, mildly relativistic regime are impossible there.

A glimpse at the workings of gravitation in a different regime is offered by binary pulsars, i.e. systems comprising of a pulsar (which allows accurate tracking of the orbital period), and another compact star (typically a neutron star or a white dwarf). These systems, the first of which was discovered in 1974~\cite{Hulse:1974eb}, have velocities that are not much larger than in the solar-system ($v\lesssim 10^{-3} c$), but present large gravitational fields/curvatures inside the compact stars. In this mildly relativistic \textit{but} strong-field regime, GR predicts that gravitational waves (GWs) should be copiously emitted, thus carrying enough energy and angular momentum away  
from the binary to produce an observable backreaction on its orbital evolution. Indeed, as the binary shrinks as a result of GW emission, its period should decrease. This effect has indeed been observed in binary pulsars, and the period's rate of change matches perfectly the GR prediction, thus providing indirect evidence of the existence of GWs~\cite{Damour:1990wz,Damour:1991rd}. 

Finally, ``advanced'' ground-based GW interferometers, such as Advanced LIGO, Advanced Virgo and KAGRA, will come online in the next few years, and are expected to detect GWs directly before the end of this decade. Because the main GW sources for these detectors are expected to be binaries of neutron stars and/or black holes at small separations (and thus with relative velocities $v\sim c$), these interferometers will provide the first test of GR in the currently unexplored highly relativistic \textit{and} strong field regime (see e.g. Refs.~\cite{Yunes:2013dva,Berti:2015itd} for two recent reviews).

Despite GR's past triumphs and the busy experimental activity to test it even further with GWs, signs that something could be wrong with our understanding of gravity might already be hidden
in plain sight in cosmological data. In the last two decades, observations of the Cosmic Microwave Background (CMB), type-Ia supernovae and the large-scale structure of the universe pointed to the existence of a Dark Matter component and a cosmological constant (or a dynamical Dark Energy component); see e.g. Ref.~\cite{Frieman:2008sn} for a review. While this ``concordance'' $\Lambda$CDM model is in agreement with essentially all observations so far, it is theoretically unappealing because ``naturalness'' arguments can explain neither the small value of the cosmological constant compared to the Planck scale, nor why it has only recently started to drive the expansion of the universe~\cite{Weinberg:1988cp,Weinberg:2000yb,Carroll:2000fy}.
In the light of the $\Lambda$CDM model's ``unattractiveness'', it makes sense at least to ask the question of whether the existence of a Dark Sector may simply be an artifact of our use of GR to explain cosmological observations. Because these observations are well within the weak-field, mildly relativistic regime tested in the solar system, the answer to this question would seem to be negative. This reasoning, however, neglects some important considerations. 

First, the Newtonian and PN dynamics that are verified in the solar system and in binary pulsars are expansions around the Minkowski geometry. 
This is not suitable for describing cosmological scales, which are rather described by the Robertson-Walker geometry (and by perturbative expansions around it). While in GR perturbative expansions around the two space-times behave in similar ways, the same is not guaranteed to happen in more general gravity theories. For instance, certain gravity theories may have a \textit{screening mechanism} built in, which triggers modifications away from the GR behavior only under certain conditions~\cite{Vainshtein:1972sx,Khoury:2003rn,Brax:2010gi}, e.g. on large cosmological scales. 
It is remarkable that hints in favor of such a screening mechanism might be hidden in already available cosmological data. Indeed, observations of velocities on galactic and galaxy-cluster scales seem to point at the existence of a universal acceleration scale $a_0 = 1.2 \times 10^{-10}$ m$/$s$^2 \sim c H_0$ (where $H_0$ is the present Hubble rate). 

The appearance of such a universal scale is not an obvious feature of the  $\Lambda$CDM model, which in order to interpret these data has to be supplemented with hypotheses about the baryonic physics
and its feedback on the growth of structures (see e.g. Refs.~\cite{2014arXiv1412.2712S,2010PhR...495...33B} for recent reviews about galaxy formation in the $\Lambda$CDM model).
Even worse, these additional assumptions need to be finely tuned to correctly reproduce the data, at least in specific cases~\cite{Famaey:2011kh,2011PhRvL.106l1303M,2012AJ....143...40M}. 
The appearance of a universal scale linked to the Hubble rate fits instead in the logic presented above, in which deviations from the GR behavior appear when one moves away from perturbative expansions over Minkowski space toward expansions over a Robertson-Walker space-time.\footnote{The Robertson-Walker geometry globally reduces to the Minkowski one when the Hubble expansion rate is zero at all times.} Alternatively, one can devise gravity theories that include an acceleration-based screening mechanism, whereby GR is recovered in high-acceleration regimes (i.e. in the solar system and binary pulsars) and modified in low-acceleration ones, where the $\Lambda$CDM postulates the existence of Dark Matter (and Dark Energy). Indeed,  the appearance of the universal acceleration $a_0$ in observations of galaxies and galaxy clusters may be a guiding principle in constructing a theory of gravity alternative to Newtonian theory/GR, in the same way in which Kepler's laws were instrumental in overcoming the Aristotelian/Ptolemaic mechanics. 
These acceleration-based attempts, which are known under the name of ``Modified Newtonian Dynamics'' (MOND)~\cite{Milgrom:1983ca,Milgrom:1983pn,Milgrom:1983zz}, 
are not yet completely successful, because to explain observations of galaxy clusters they still need 
some residual ``dark missing baryons'', with mass roughly twice that of observed baryons~\cite{Famaey:2011kh} and possibly in the form of molecular hydrogen~\cite{2009A&A...496..659T}. [Note
that this is not in contrast with the estimate of the baryon density coming from  Big-Bang Nucleosynthesis (BBN), since about 30\% of the baryons 
produced during BBN are still undetected, and only 4\% are observed  in clusters~\cite{2012ApJ...759...23S}.] Nevertheless, the appearance of a universal scale in the data is a genuine \textit{empirical} 
feature, the explanation of which is still poorly understood.

Another independent motivation for considering possible modifications of GR comes from its intrinsic incompatibility with quantum field theory, i.e. the long-known fact that GR, when quantized, is not power-counting renormalizable in the ultraviolet (UV) regime, where it should be replaced by a (yet unknown) quantum theory of gravity. In addition, GR generically predicts the existence of curvature singularities in time evolutions starting from regular initial data. Even though these singularities are conjectured to be always enclosed by black-hole horizons and thus inaccessible to outside observers~\cite{Wald:1997wa,Penrose:1999vj}, their existence is a disturbing feature that one expects should be solved by a full quantum theory of gravity.

A candidate quantum-gravity theory that addresses these two problems is given by Ho\v rava gravity~\cite{Horava:2009uw,Blas:2009qj}. This theory breaks boost-symmetry 
(and thus Lorentz invariance) in the gravitational sector
by adding to the action terms that are of fourth and sixth order in the \textit{spatial} derivatives of the metric. In simpler scalar toy models, these terms are enough to
achieve UV power counting renormalizability~\cite{Horava:2009uw,Visser:2009fg}, and the hope is that the same will happen for spin-2 gravitons. Also, the presence of the
higher-order terms in the spatial derivatives is expected to smooth the curvature singularities typically forming in GR evolutions~\cite{Blas:2014aca}. 
On astrophysical scales, Ho\v rava gravity is practically indistinguishable from its low-energy limit, sometimes called ``khronometric theory''~\cite{Blas:2009qj,Blas:2010hb}. This theory has been extensively studied, thanks also to the fact that it is closely related~\cite{Jacobson:2010mx,Jacobson:2013xta} to another previously introduced and actively scrutinized family of phenomenological boost-violating gravity theories, i.e. Einstein-\AE ther theories~\cite{Jacobson:2000xp,Jacobson:2008aj}. Remarkably, khronometric theory (and thus Ho\v rava gravity) has been
shown to pass all experimental tests, i.e. submillimeter tests~\cite{Blas:2014aca}, absence of gravitational \v Cerenkov radiation~\cite{Elliott:2005va}, solar-system experiments~\cite{Foster:2005dk,Blas:2010hb,Blas:2011zd}, binary- and isolated-pulsar observations~\cite{Yagi:2013ava,Yagi:2013qpa}, and existence of regular black holes forming from gravitational collapse (so as to agree with astrophysical observations of black-hole candidates)~\cite{Garfinkle:2007bk,Eling:2006ec,Barausse:2011pu,Blas:2011ni,Barausse:2012ny,Barausse:2012qh,Barausse:2013nwa}, in regions of parameter space where khronometric theory is stable at both the classical and quantum levels.

An attempt at modifying khronometric theory and Ho\v rava gravity to account for the presence of a universal acceleration scale in galaxy and galaxy-cluster data was done in Ref.~\cite{Blanchet:2011wv}, which introduced a theory that reduces to a (very special) khronometric/Ho\v rava-gravity theory in high-acceleration regimes, and which produces a MOND behavior in the low-acceleration, non-relativistic/weak-field regime relevant for galaxies and clusters. This theory clearly shares both the flaws and the blessings of MOND that we mentioned above, namely it accounts for the appearance of a universal acceleration without finely tuned baryonic physics/feedback, but may \textit{still} need some form of Dark Matter in the center of galaxy clusters.\footnote{The necessary amount of Dark Matter is smaller than in the
$\Lambda$CDM model. Indeed, as mentioned above, it might be sufficient to identify this Dark Matter with some of the ``missing dark baryons''
that are predicted by BBN, but which are not observed in the local universe in the form of visible matter. It has been proposed that these dark baryons may be in the form of molecular hydrogen~\cite{2009A&A...496..659T}.}
Also, the theory of Ref.~\cite{Blanchet:2011wv} is related to some of the older theories
proposed to obtain a MOND-like phenomenology in the non-relativistic limit  -- namely tensor-vector-scalar gravity  (TeVeS)~\cite{Bekenstein:2004ne} and generalized Einstein-\AE ther theories~\cite{Zlosnik:2006sb,Zlosnik:2006zu}; c.f. also
Ref.~\cite{Famaey:2011kh} for an extensive review of the theories giving a MOND phenomenology --, but is better motivated theoretically,
because it reduces to a viable quantum gravity model such as Ho\v rava gravity at high accelerations.

In this paper, we will work out the 1PN expansion of the theory of Ref.~\cite{Blanchet:2011wv}, in both the high- and low-acceleration regimes. We will show that
\textit{if} one imposes that the theory reduces to GR in the high-acceleration regime,  a \textit{strong-coupling problem} arises in the low-acceleration regime when 1PN terms are considered in the dynamics, and this would ruin the agreement with the observed rotation curves of galaxies. Indeed, we will show that while these observations are reproduced in
the Newtonian limit, the 1PN dynamics is strongly coupled, as a result of which the 1PN terms become dominant over the Newtonian ones in regimes accessible by galaxy rotation curves.
However, we will then show that a simple slight generalization of the theory of Ref.~\cite{Blanchet:2011wv} allows us to avoid this strong-coupling problem, i.e. one can obtain a fully viable theory by relaxing the assumption that the dynamics should reduce \textit{exactly} to GR in the high-acceleration limit.
We will therefore end up with a theory that \textit{(i)} presents a well-behaved (i.e. perturbative) PN expansion at all accelerations;
\textit{(ii)} passes submillimeter, pulsar and solar-system tests; \textit{(iii)} reduces to a general khronometric theory (and thus to Ho\v rava gravity) at high accelerations; \textit{(iv)} gives a MOND-like phenomenology at the low accelerations characterizing galaxies and clusters.

This paper is organized as follows. In section \ref{sec:Einstein-Aether and khronometric theories} we introduce the theories under investigation. The dynamics of these theories
in the high-acceleration regime, as well
as the experimental/theoretical constraints on it, are discussed in section \ref{sec:Constraints}.
The low-acceleration regime, and in particular the 1PN dynamics, is discussed in section \ref{sec:khronometric-MOND theory}, both
in the general case and for the special case of a galaxy accreting gas. We show
that the low-acceleration 1PN dynamics is strongly coupled in a certain region of parameter space, and that this may
jeopardize the agreement of the theory with data on the scales of galaxies. In section \ref{sec:A new constraint}
we identify this region, and show that the theories that we consider remain viable in large portions of
the parameter space. A final discussion is then presented in section \ref{sec:discussion}.

 We will also use a metric signature $(-+++)$, and we will denote space-time indices by Greek letters and
spatial ones by Latin letters.  Spatial vectors are also denoted by an over-arrow.
We will set $c=1$ throughout this paper, except when dealing with PN expansions in sections \ref{subsec:Post-Newtonian expansion}, \ref{sec:Simplified physical system}
and in the Appendix, where we reinstate the factors $1/c$ as PN book-keeping
parameters. We will denote in particular the  $n/2$-th PN order by $O(n)$, i.e. $O(n)\equiv O(c^{-n})$.

\section{Khronometric theories with a MOND non-relativistic limit}
\label{sec:Einstein-Aether and khronometric theories}

The action of Ho\v rava gravity~\cite{Horava:2009uw,Blas:2009qj} can be written as
\begin{multline}
\label{eq:action-H-full}
S_{H}= \frac{1-\beta}{16\pi G}\int dT d^3x \, N\sqrt{\gamma}\left(L_{\rm kh}+\frac{L_4}{M_\star^2}+\frac{L_6}{M_\star^4}\right)\\+ S_{\rm m}(\bm{\varphi},g_{\mu\nu})\,,
\end{multline}
where the spacetime has been foliated in spacelike hypersurfaces, and the metric $g_{\mu\nu}$ has been accordingly decomposed in 3+1 form, i.e. 
we introduce the lapse function $N=(-g^{00})^{-1/2}$, the shift 3-vector $N_i=g_{0i}$, the induced 3-metric $\gamma_{ij}=g_{ij}$ (as well as its determinant $\gamma$) 
and the extrinsic curvature 
\begin{equation}
K_{ij}=\frac{1}{2N}(\partial_t \gamma_{ij}-D_i N_j-D_j N_i)\,,
\end{equation}
with $D_i$ denoting covariant derivatives relative to the geometry of the spacelike hypersurfaces (i.e. $D_i \gamma_{jk} = 0$). 
The
matter part of the action is instead represented by $S_{\rm m}$, where the matter fields $\bm{\varphi}$ couple to the covariant four-dimensional metric $g_{\mu\nu}$, so as to enforce the weak equivalence principle and to confine Lorentz violations in the gravitational sector (at tree level)~\cite{Liberati:2012jf}. 
The Lagrangian density $L_{\rm kh}$
is the most generic one at quadratic order in derivatives (up to total divergences), i.e.  
\begin{equation}\label{Lkh} L_{\rm kh}=K_{ij}K^{ij} - \frac{1+\lambda}{1-\beta} K^2
\\+  \frac{1}{1-\beta}{}^{(3)}\!R +  \frac{\alpha}{1-\beta}\, a_ia^i\,,
\end{equation}
where $a_i = \partial_i \ln N$, $K=\gamma^{ij} K_{ij}$ is the trace of the extrinsic curvature, and $\alpha$, $\beta$ and $\lambda$ are dimensionless free parameters.
The parameter $\alpha$ regulates (among other things) the relation between the 
``bare'' gravitational constant $G$ appearing in the action and the ``Newtonian'' gravitational constant $G_N$ 
measured by a Cavendish experiments, which
turns out to be
\begin{equation}\label{GN}
G_N=\frac{2G}{2-\alpha}\,.
\end{equation}

The $L_4$ and $L_6$ Lagrangian densities are instead of fourth- and sixth-order, respectively,  in the \textit{spatial} derivatives $D_i$, but contain no 
time derivatives~\cite{Horava:2009uw,Blas:2009qj,Sotiriou:2009bx}. This ensures that the theory does not suffer from the Ostrogradski instability~\cite{ostrogradski}\footnote{See also section 2 of Ref.~\cite{Woodard:2006nt} for a pedagogical review of the Ostrogradski instability.}, and most of all provides the anisotropic scaling necessary for power-counting renormalizability~\cite{Horava:2009uw,Visser:2009fg}. For dimensional reasons, the $L_4$ and $L_6$ terms must be suppressed by an energy scale $M_{\star}$.
This scale must be $M_{\star}\lesssim 10^{16}$ GeV to ensure that the theory remains perturbative at all scales, which is a necessary condition for power-counting renormalizability arguments to apply. 
Also, experimental constraints put lower bounds on $M_\star$. More precisely, to ensure agreement with submillimeter experiments~\cite{Adelberger:2006dh,Kapner:2006si}, 
it must be $M_{\star}\gtrsim 10^{-2}$ eV, and even more stringent bounds may be possible depending on the details of the percolation of the Lorentz violations in the matter sector beyond tree level.
Indeed, observations of the synchrotron emission from the Crab Nebula show that this percolation
should be suppressed if the theory is to remain viable and perturbative on all scales~\cite{Liberati:2012jf}. Several mechanisms have been proposed to suppress the percolation of
Lorentz violations from the gravity sector into the matter one, including fine tuning, ``gravitational confinement''~\cite{Pospelov:2010mp}, ``custodial symmetries'' 
(e.g. softly broken supersymmetry~\cite{GrootNibbelink:2004za,Pujolas:2011sk}), or dynamical emergence of Lorentz symmetry at low energies in the matter sector, e.g. due to renormalization 
group flows~\cite{Chadha:1982qq,Bednik:2013nxa}. We refer the reader to
Ref.~\cite{Liberati:2013xla} for a review of these possibilities, and assume in this paper that one of these mechanisms suppresses the percolation to acceptable levels, so that
the bound on $M_\star$ is $M_{\rm obs}\lesssim M_\star\lesssim 10^{16}$ GeV, with $M_{\rm obs}\gtrsim 10^{-2}$ eV.
For these values of $M_\star$ and at the low energies typically characterizing astrophysical observations, the higher-order terms $L_4$ and $L_6$ are typically negligible~\cite{Barausse:2013nwa}, with the possible exception of black holes (whose causal structure does depend on the presence of the $L_4$ and $L_6$ terms, c.f. the concept of universal horizon~\cite{Barausse:2011pu,Blas:2011ni}). When those terms are neglected, Ho\v rava gravity coincides with ``khronometric'' theory~\cite{Horava:2009uw,Blas:2009qj}, i.e. a theory with the action \eqref{eq:action-H-full}, but with
$L_4$ and $L_6$ set to zero \textit{ab initio}.

A useful way of writing the action of khronometric theory is to introduce a scalar field $T$ (the ``khronon'') defining the 3+1 foliation, i.e. such 
that the constant-$T$ surfaces coincide with the foliation's spacelike hypersurfaces. Because of this requirement, this scalar field must have a
timelike gradient, i.e. $g^{\mu\nu} \partial_\mu T \partial_\nu T<0$ within our conventions. In terms of this khronon field, the action of
khronometric theory [i.e. Eq.~\eqref{eq:action-H-full} with $L_4=L_6=0$] can be written in covariant form as~\cite{Jacobson:2013xta,Jacobson:2010mx}
\begin{multline}\label{eq:khrono-action-covariant}
S = \dfrac{1}{16\pi G}\int \!\!d^4 x \sqrt{-g} \Bigl[R - \frac13 (\beta+3\lambda) \theta^2\\-\beta \sigma_{\mu\nu} \sigma^{\mu\nu}+\alpha a_\mu a^\mu\Bigr]
+ S_\textup{mat}(\bm{\varphi},g_{\mu\nu}),
\end{multline}
where $g$ is the metric's determinant, $R$ is the (four-dimensional) Ricci scalar, 
\begin{equation}\label{eq:gradient of scalar field}
n_{\mu} =  -\dfrac{\partial_{\mu}T}{\sqrt{-g^{\alpha\beta}\partial_{\alpha}T \partial_{\beta}T}}
\end{equation}
is the (timelike) unit-norm vector field orthogonal to the foliation, and
\begin{gather}
a^\mu= n^\nu\nabla_\nu n^\mu\,\\ 
\theta=\nabla_\mu n^\mu\,\\
\sigma_{\mu\nu}=\nabla_{(\nu}n_{\mu)}+a_{(\mu}n_{\nu)}-\frac13\theta \gamma_{\mu\nu} 
\end{gather}
(with $\gamma_{\mu\nu}=g_{\mu\nu}+n_\mu n_\nu$ the projector onto the spacelike hypersurfaces)
are the acceleration, expansion and shear of the congruence defined by $n_\mu$, i.e.
$\nabla_\mu n_\nu= - a_\nu n_\mu+\sigma_{\mu\nu}+\frac13 \theta \gamma_{\mu\nu}$.
[Note that the vorticity $\omega_{\mu\nu}=\nabla_{[\nu}n_{\mu]}+a_{[\mu}n_{\nu]}=\partial_{[\nu}n_{\mu]}+a_{[\mu}n_{\nu]}$
vanishes identically because of  Eq.~\eqref{eq:gradient of scalar field}.]

It should be noted that the action \eqref{eq:khrono-action-covariant} is very similar to that of Einstein-\AE ther theory~\cite{Jacobson:2000xp,Jacobson:2008aj}, with the caveat that in that theory
the vector $n_\mu$ is assumed to be timelike and unit-norm (thus $n^\mu n_\mu=-1$) but  \textit{not}  hypersurface orthogonal, i.e.  $n_\mu$ is a full-fledged (timelike and unit-norm) vector that \textit{cannot} be expressed in terms of a scalar through Eq.~\eqref{eq:gradient of scalar field} at the level of the action. For this reason, 
 the vorticity of $n_\mu$ is not zero, and the most generic action for Einstein-\AE ther theory is obtained by adding 
to the action \eqref{eq:khrono-action-covariant}
an extra term $c_\omega \omega_{\mu\nu} \omega^{\mu\nu}$ ($c_\omega$ being a dimensionless coupling constant), as well as a term $\xi (n_\mu n^\mu+1)$ (where $\xi$ is a Lagrange multiplier) enforcing the unit-norm timelike character of the vector field $n_\mu$. 

Reference~\cite{Blanchet:2011wv} proposed to modify the action of khronometric theory at the very large scales (i.e. very low energies) characterizing cosmological
observations, i.e. in the infrared limit. The idea, as we outlined in the introduction, is that the cosmological evidence for Dark Matter 
comes from systems with accelerations $a<a_0 \approx H_0/6$, and the theory introduced in Ref.~\cite{Blanchet:2011wv} seeks to reproduce the Dark-Matter phenomenology 
without any actual Dark Matter (with the possible exception, as explained above, of some ``dark baryons'' on  galaxy-cluster scales) by modifying the
gravity theory in that low-acceleration regime. This corresponds to modifying the gravity theory on cosmological scales $\gtrsim 1/a_0$, or equivalently energies
$\lesssim \hbar \,a_0 \sim 10^{-34} $ eV.
More precisely, Ref.~\cite{Blanchet:2011wv} considered a modified
khronometric theory
with action
\begin{equation}
S = \dfrac{1}{16\pi G}\int \!\!d^4 x \sqrt{-g} \Bigl[R + f(a)\Bigr]+ S_\textup{mat}(\bm{\varphi},g_{\mu\nu}),
\end{equation}
with $a=\sqrt{\gamma_{\mu\nu} a^\mu a^\nu}$ and $n_\mu$ still given by Eq.~\eqref{eq:gradient of scalar field}. Reference~\cite{Blanchet:2011wv} then showed that in order to obtain
a MOND-like phenomenology in the non-relativistic, low-acceleration limit, the 
free function $f(a)$ must asymptote to $f(a)\approx -2\Lambda_0+2a^2-4 a^3/(3 a_0)$ (where $\Lambda_0$ is a constant) for $a\ll a_0$, while they propose
the limit $f(a)\sim -2\Lambda_{\rm obs}$ ($\Lambda_{\rm obs}$ being the measured cosmological constant) for $a\gg a_0$ in order to reproduce GR (with a 
cosmological constant) in the high-acceleration regime. As we will show below, however, this theory does \textit{not} produce
a perturbative post-Newtonian (PN) expansion in time-dependent situations such as those of interest for cosmology and astrophysics, i.e. the PN expansion
turns out to be strongly coupled. We will show, however, that this problem can be avoided with a slight modification of the theory of Ref.~\cite{Blanchet:2011wv}, namely one
with action
\begin{multline}\label{action-full-our-theory-covariant}
S = \dfrac{1}{16\pi G}\int \!\!d^4 x \sqrt{-g} \Big[R - \frac13 (\beta+3\lambda) \theta^2-\beta \sigma_{\mu\nu} \sigma^{\mu\nu}\\+ f(a)\Big]+ S_\textup{mat}(\bm{\varphi},g_{\mu\nu}),
\end{multline}
where again $a=\sqrt{\gamma_{\mu\nu} a^\mu a^\nu}$, $n_\mu$ is given by Eq.~\eqref{eq:gradient of scalar field}, and $f(a)$ satisfies again the asymptotic limit
$f(a)\approx -2\Lambda_0+2a^2-4 a^3/(3 a_0)$ for $a\ll a_0$. Note that this action can be rewritten in a 3+1 foliation adapted to the khronon, in the same way in which khronometric theory can be written in the two equivalent forms \eqref{eq:action-H-full}-\eqref{Lkh} and \eqref{eq:khrono-action-covariant}, thus obtaining
\begin{multline}
\label{action-full-our-theory-adapted}
S_{H}= \frac{1-\beta}{16\pi G}\int dT d^3x \, N\sqrt{\gamma}\Big(K_{ij}K^{ij} - \frac{1+\lambda}{1-\beta} K^2
\\+  \frac{1}{1-\beta}{}^{(3)}\!R +  \frac{f(a)}{1-\beta}\, \Big)+ S_{\rm m}(\bm{\varphi},g_{\mu\nu})\,,
\end{multline}
where $a=\sqrt{\gamma_{\mu\nu}a^\mu a^\nu}=\sqrt{\gamma_{ij}a^i a^j}$ in 3+1 form. In the high-acceleration regime relevant for 
astrophysical and experimental tests (i.e. submillimeter, solar-system and pulsar ones), 
we impose that the theory reduces to khronometric gravity (plus a cosmological constant), i.e. for $a\gg a_0$ (but $a\ll M_\star$) we choose $f(a)\sim -2 \Lambda + \alpha a^2$~\footnote{
$\Lambda$ is related to the measured cosmological constant $\Lambda_{\rm obs}$ by $\Lambda=\Lambda_{\rm obs} G/G_c$, where
$G_c=2 G/(2 + \beta + 3 \lambda)$ is the gravitational constant appearing in the Friedmann equations~\cite{Blas:2012vn}.
In pratice, since BBN metal abundances and binary-pulsar observations constrain $|\beta,\lambda|\lesssim$ a few $\times\, 0.01$~\cite{Yagi:2013ava,Yagi:2013qpa}
 (c.f. also section~\ref{sec:Constraints}), it must be $\Lambda\sim\Lambda_{\rm obs}$.
}, while
at higher energies (i.e. $a\gg M_\star$) we may identify our theory with the full Ho\v rava theory.

Of course, it remains to be seen whether the renormalization-group flow is compatible with this choice for the coupling function $f(a)$, i.e.
whether the MOND-like theory of Ref.~\cite{Blanchet:2011wv} (or a similar one, c.f. discussion in section \ref{sec:discussion}) is an infrared fixed point of the 
renormalization-group flow of Ho\v rava gravity.
From this point of view, our treatment is purely phenomenological.

\section{The high-acceleration regime}
\label{sec:Constraints}
As discussed above, for high accelerations (i.e. high energies) $a\gg a_0$, our theory reduces to Ho\v rava gravity. 
In particular, for the accelerations $a_0\ll a\ll M_\star$ relevant for experiments on Earth and in the solar-system,
as well as for most astrophysical (non-cosmological) observations, the theory described by actions \eqref{action-full-our-theory-covariant} or
\eqref{action-full-our-theory-adapted} reduces to khronometric theory. Here, we therefore review the experimental
constraints on the coupling constants $\alpha$, $\beta$ and $\lambda$ of khronometric theory. Clearly, those constraints also apply to our theory.

A linear expansion of the field equations of khronometric theory on a Minkowski background shows that the theory presents
a spin-2 graviton polarization propagating with speed $c_t$, as well as a spin-0 one with propagation speed $c_s$. These speeds
are given by~\cite{Blas:2010hb,Blas:2011zd}
\begin{gather}\label{eq:stability}
c_t^2=\frac{1}{1-\beta} \,,\\ 
c_s^2=\frac{(\alpha-2)(\beta+\lambda)}{\alpha(\beta-1)(2+\beta+3\lambda)} \,.
\end{gather}
To avoid gradient instabilities on Minkowski space, one must impose $c_s^2>0$ and $c_t^2>0$. These conditions also ensure that energies are positive~\cite{Blas:2010hb,Garfinkle:2011iw}, thus avoiding ghost instabilities. Even more stringently,
to prevent ultra-high energy cosmic rays from losing energy to gravitons by vacuum \v Cerenkov radiation~\cite{Elliott:2005va}, the gravitational
modes must also propagate luminally or superluminally, i.e. $c_t^2\geq1$ and $c_s^2\geq1$.

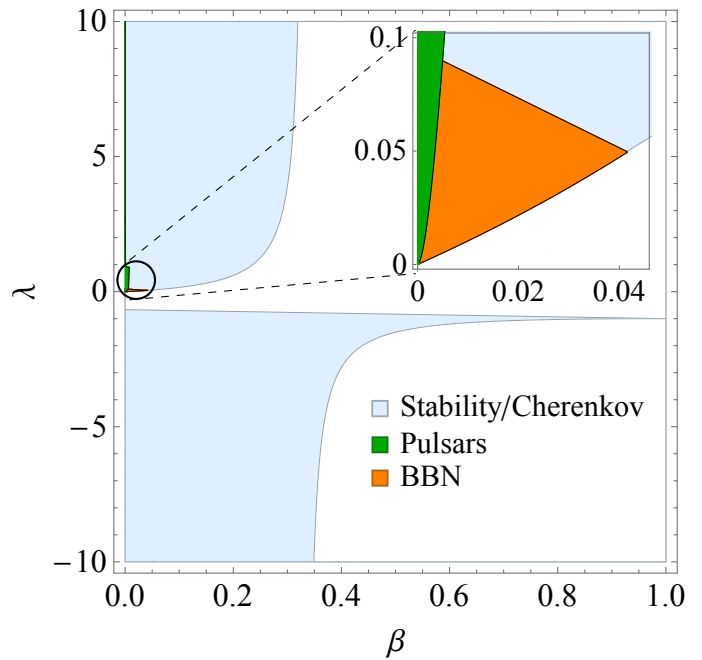
\begin{figure}[tb]
\centering
\begin{tikzpicture}
    \node[anchor=south west,inner sep=0] (image) at (0,0) {\includegraphics[scale=0.57]{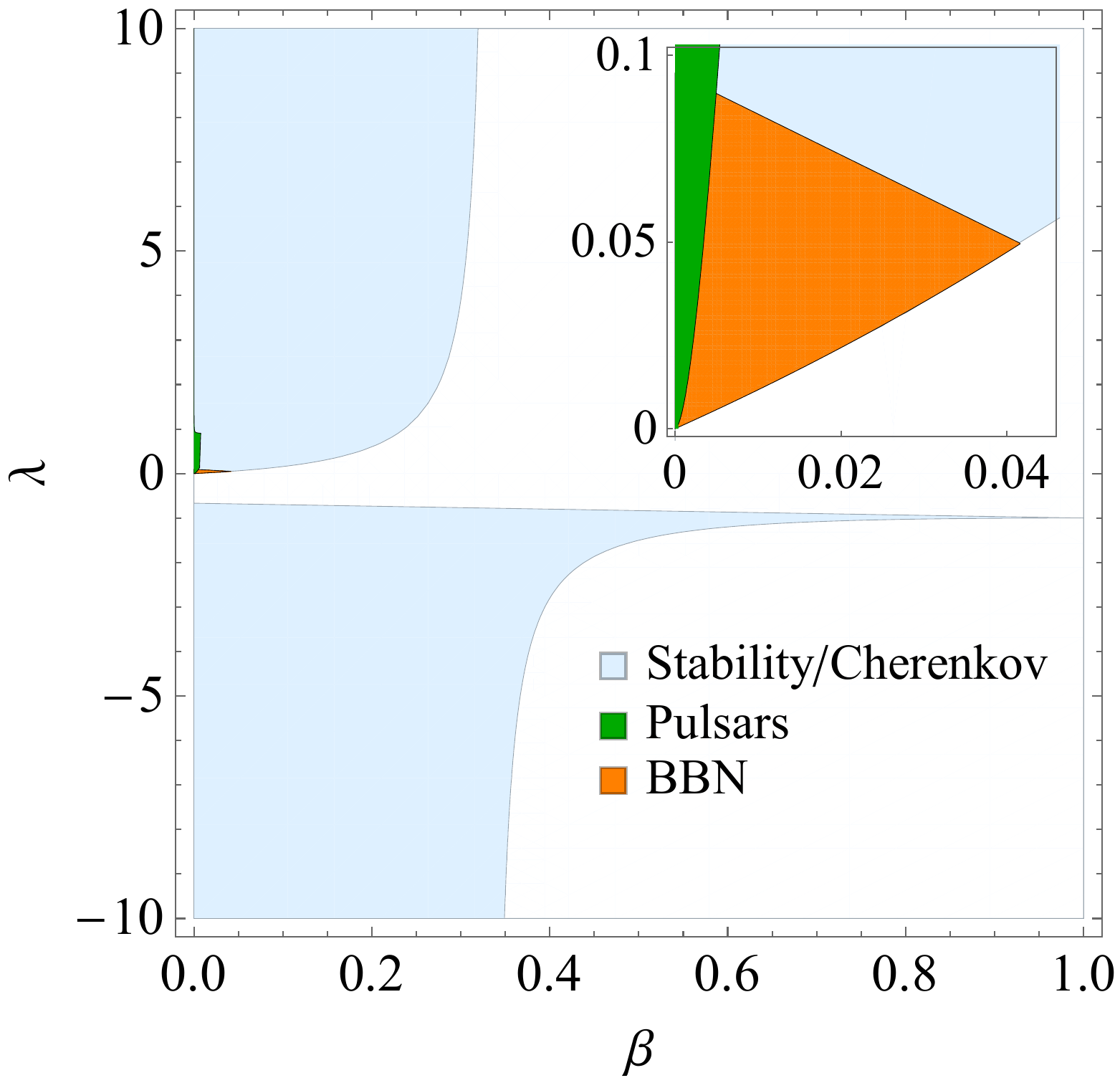}};
    \begin{scope}[x={(image.south east)},y={(image.north west)}]
        \draw[black,thick, radius=0.25cm] (0.19,0.58) circle ;
        \draw[black,dashed] (0.18,0.61) -- (0.60,0.96) ;
        \draw[black,dashed] (0.18,0.55) -- (0.60,0.59) ;
        \draw[white,ultra thick] (0.585,0.963) -- (0.95,0.963) ;
        
    \end{scope}
\end{tikzpicture}
\caption{Allowed parameter space for $\beta$ and $\lambda$~\cite{Yagi:2013ava,Yagi:2013qpa}, from observations with
accelerations $a\gg a_0$ (solar-system tests, absence of \v Cerenkov vacuum radiation, BBN, pulsars) and from stability requirements (no gradient/ghost instabilities).}
\label{fig:vincoli} 
\end{figure}

To ensure that khronometric theory agrees with experiments at the level of the solar-system, one can solve the field
equations at first PN order, and compute the Parametrized PN (PPN) parameters~\cite{Will:1993ns}.
All these parameters turn out to be the same as in GR, with the exception of the preferred-frame parameters $\alpha_1$ and $\alpha_2$~\cite{Blas:2010hb,Blas:2011zd}
\begin{align}\label{eq:alpha1}
\alpha_1&=\dfrac{4(\alpha - 2\beta)}{\beta-1}\,,\\\label{eq:alpha2}
\alpha_2&=\dfrac{(\alpha-2\beta)[-\beta(3+\beta+3\lambda) - \lambda +\alpha(1+\beta+ 2\lambda)]}{(\alpha-2)(\beta-1)(\beta+\lambda)} \,.
\end{align}
Solar-system tests constrain $\vert\alpha_1\vert \lesssim 10^{-4}$ and $\vert\alpha_2\vert \lesssim 10^{-7}$~\cite{Will:2014xja}. To satisfy these 
bounds, one can simply impose $\alpha=2\beta+{\cal O}(\alpha_1,\alpha_2)$ at leading order in $\alpha_1$ and $\alpha_2$. This is sufficient to satisfy the constraints
on both $\alpha_1$ and $\alpha_2$, since both quantities are proportional to the combination $\alpha - 2\beta$. This allows decreasing the
dimension of the theory's parameter space from three (i.e. $\alpha,\beta,\lambda$) to two (i.e. $\beta,\lambda$).\footnote{One may also impose the bounds $\vert\alpha_1\vert \lesssim 10^{-4}$ and $\vert\alpha_2\vert \lesssim 10^{-7}$ without exploiting the fact that both  $\alpha_1$ and $\alpha_2$ are proportional to $\alpha - 2\beta$. However, this
  would produce a one-dimensional parameter space, which turns out to be a subset of the 
two-dimensional parameter space that one obtains by choosing  $\alpha=2\beta+{\cal O}(\alpha_1,\alpha_2)$.
In this sense one may choose $\alpha\approx2\beta$ without loss of generality, c.f. Ref.~\cite{Yagi:2013ava} for a detailed discussion.}

Once the constraints discussed above are accounted for, the viable parameter space $(\beta,\lambda)$ 
is given by the cyan region in Fig.~\ref{fig:vincoli}. Additional bounds on the parameters then come, as mentioned above, from the requirement that
BBN produce the observed element abundances~\cite{Blas:2012vn,Carroll:2004ai,Yagi:2013ava,Yagi:2013qpa} 
(orange region in Fig.~\ref{fig:vincoli}). Also, stringent bounds (represented in green in
Fig.~\ref{fig:vincoli}) come from
the absence of any anomalous precession in observations of isolated pulsars~\cite{Yagi:2013ava,Yagi:2013qpa}, as well as from the
change of the measured period of binary pulsars under gravitational-wave emission~\cite{Yagi:2013ava,Yagi:2013qpa}. Indeed, the latter effect
puts very strong constraints on $\beta$ and $\lambda$, because the presence of a khronon field
coupled non-minimally to the metric causes the appearance of dipolar fluxes in the 
gravitational-wave emission from binary systems, besides the quadrupolar fluxes of GR~\cite{Yagi:2013ava,Yagi:2013qpa}. Because 
binary-pulsar observations are in good agreement with the GR predictions, these dipolar fluxes must be suppressed
by sufficiently small values of the coupling constants. 

Nevertheless, as is clear from Fig.~\ref{fig:vincoli},
there is a sizeable region of parameter space 
where khronometric theory [and thus
the theory described by Eqs.~\eqref{action-full-our-theory-covariant} or
\eqref{action-full-our-theory-adapted}] is viable, 
around the limit $\beta=\lambda=0$ (in which GR is recovered at high accelerations). 
Note that in this viable region of parameter space, black-hole
solutions that arise from gravitational collapse~\cite{Garfinkle:2007bk} have also been shown to exist~\cite{Eling:2006ec,Barausse:2011pu,Blas:2011ni,Barausse:2012ny,Barausse:2012qh,Barausse:2013nwa}. These solutions present
properties compatible with current electromagnetic observations of black-hole candidates (i.e. their exterior geometry is very close
to the black-hole solutions of GR)~\cite{Barausse:2011pu,Barausse:2013nwa}.

Finally, as discussed in the previous section, 
our theory reduces to Ho\v rava gravity in the UV regime $a\gg M_\star$. Therefore, constraints coming from
sub-millimeter tests of the $1/r^2$ decay of the Newtonian attraction force are satisfied provided that $M_\star\gtrsim 10^{-2}$ eV~\cite{Blas:2014aca}, while
tests of Lorentz invariance in the matter sector will be passed provided that a suitable mechanism exists that suppresses the percolation of
Lorentz violations from gravity to the matter sector (c.f. discussion and references above). In addition, as alluded above, the higher-order derivative terms
of Ho\v rava gravity are important for the propagation of signals in a black-hole spacetime, but are not expected to destroy its causal structure (which still possesses a universal horizon from which no signals can escape, not even with infinite propagation speed~\cite{Barausse:2011pu,Blas:2011ni}).

\section{The low-acceleration regime}
\label{sec:khronometric-MOND theory}

In this section, we will study the 1PN expansion of the theory described by actions \eqref{action-full-our-theory-covariant} or
\eqref{action-full-our-theory-adapted}. While our treatment is
valid in both the high- and low-acceleration regimes, we will focus mostly on the latter. Indeed, as discussed in the previous section, at high accelerations
the theory reduces to khronometric theory/Ho\v rava gravity, for which the 1PN expansion has already been derived in Refs.~\cite{Blas:2010hb,Blas:2011zd}, and shown to agree
with experimental constraints coming from solar-system tests in large portions of the parameter space. As a check of our calculation, we will
however verify that we reproduce the 1PN expansion of khronometric theory derived in Refs.~\cite{Blas:2010hb,Blas:2011zd}, confirming in particular their expressions for the preferred-frame
parameters $\alpha_1$ and $\alpha_2$ [Eqs.~\eqref{eq:alpha1} and \eqref{eq:alpha2}]. 

The calculation of the 1PN expansion in the low-acceleration regime, which we present below, 
may \textit{a priori} be expected to be of purely academic interest. After all, the tests of the PN dynamics of GR (solar-system tests and
binary pulsars) are in high-acceleration regimes, while systems with accelerations $a\ll a_0$ (such as those encountered in cosmology)
have velocities too small relative to the speed of light to test even the first PN order (with the accuracy of current data). 

Nevertheless, we will show that surprises arise in the course of the calculation. In particular, we will show that if one sets
the couplings $\beta$ and $\lambda$ to zero (as in the original theory of Ref.~\cite{Blanchet:2011wv}) or to values below a certain threshold, 
the 1PN expansion in the low-acceleration regime becomes strongly coupled. We will show that this 
prevents the theory from reproducing the Dark-Matter phenomenology at accelerations $a\ll a_0$, at least in dynamical/time-dependent situations and
unless the couplings $\beta$ and $\lambda$ are significantly different from zero. We will show this explicitly by calculating the rotation
curves of the gas surrounding a galaxy (whose mass grows due to accretion from the intergalactic medium -- IGM) at 1PN order. Based on this toy model, we will
then compute a lower bound on the combination $|\lambda+\beta|$ [c.f. Eq.~\eqref{eq:bound}], i.e. we will determine the minimum value of this combination
for which the theory avoids the aforementioned strong-coupling problem, and can thus reproduce the Dark-Matter phenomenology at low accelerations. We will show that by combining this bound with existing constraints on the couplings from the high-acceleration regime the theory remains viable in a non-negligible region of parameter 
space (c.f. Fig.~\ref{fig:vincoli_io}). 

\subsection{Modified field equations}
\label{subsec:Modified field equations}

As a first step toward computing the 1PN expansion, let us first derive the field equations by varying the action in adapted coordinates, 
i.e. Eq.~\eqref{action-full-our-theory-adapted}. The variation with respect to the lapse $N$ gives
\begin{align}\label{eq:hamiltonian_constraint}
\dfrac{^{\mbox{\tiny{(3)}}}R}{1-\beta} - K^{ij}&K_{ij} + \dfrac{1+\lambda}{1-\beta}K^2 + \dfrac{f(a)}{1-\beta}\notag\\
 &- \dfrac{2}{1-\beta}\chi a^2 - \dfrac{2}{1-\beta}D_i(\chi a^i) = \dfrac{16\pi G \mathcal{E}}{(1-\beta)c^4}\,;
\end{align}
the variation with respect to the shift $N_i$ gives  
\begin{equation}\label{eq:momentum_constraint}
D_j \left(K^{ij} - \dfrac{1+\lambda}{1-\beta}\gamma^{ij}K\right) = -\dfrac{8\pi G \mathcal{J}^i}{(1-\beta)c^4}\,;
\end{equation}
and the variation with respect to the 3-metric $\gamma_{ij}$ yields
\begin{align}\label{eq:evolution_equation}
&\dfrac{1}{1-\beta}\left(^{\mbox{\tiny{(3)}}}R^{ij} - \frac{1}{2} {^{\mbox{\tiny{(3)}}}R}\gamma^{ij} \right) + \dfrac{1}{N}D_t \left(K^{ij} - \frac{1+\lambda}{1-\beta}\gamma^{ij}K\right)\notag\\
& + \dfrac{2}{N}D_k \left(N^{(i}(K^{j)k} - K \frac{1+\lambda}{1-\beta}\gamma^{j)k})\right) + 2K^{ik}K_{k}^{j}\notag\\
& - \dfrac{1 + \beta+ 2\lambda}{1-\beta}K^{ij}K - \dfrac{1}{2}\gamma^{ij}\left(K^{kl}K_{kl} + \frac{1+\lambda}{1-\beta}K^2\right)\notag\\
& - \dfrac{1}{(1-\beta)N}\left(D^iD^j N - \gamma^{ij} D_k D^k N\right)\notag\\
& + \dfrac{1}{1-\beta}\chi a^i a^j - \dfrac{f(a)}{2(1-\beta)}\gamma^{ij} = \dfrac{8\pi G }{(1-\beta)c^4}\mathcal{T}^{ij}\,.
\end{align}
In these equations, $\chi = f'(a)/(2a)$,  $D_i$ denotes the covariant derivative compatible with $\gamma_{ij}$, while $D_t$ is a shortcut for $\partial_t - N_kD^k$. Also,
the terms $\mathcal{E}, \mathcal{J}^i, \mathcal{T}^{ij}$ come from the variation of the matter action, i.e.
		\begin{align}
		& \mathcal{E} = -\dfrac{1}{\sqrt{\gamma}}\dfrac{\delta  S_\textup{mat}}{\delta N}\label{eq:E}\,,\\
		& \mathcal{J}^i = \dfrac{1}{\sqrt{\gamma}}\dfrac{\delta  S_\textup{mat}}{\delta N_i}\label{eq:J}\,,\\
		& \mathcal{T}^{ij} = \dfrac{2}{N\sqrt{\gamma}}\dfrac{\delta  S_\textup{mat}}{\delta \gamma_{ij}}\label{eq:T}\,,
		\end{align}
		and are related to the canonical stress-energy tensor components, $T^{\mu\nu} = \left(2/\sqrt{-g}\right)\left(\delta S_\textup{mat}/\delta g_{\mu\nu}\right)$, by
		\begin{align}
		& \mathcal{E} = N^2 T^{00}\,,\\
		& \mathcal{J}^i = N(T^{0i} + N^i T^{00})\,,\\
		& \mathcal{T}^{ij} = T^{ij} - N^i N^j T^{00}\,.
		\end{align}
Note also that by combining Eq.~\eqref{eq:momentum_constraint} with the trace of Eq.~\eqref{eq:evolution_equation} (obtained by contracting that equation with $\gamma_{ij}$), we obtain 
\begin{align}\label{eq:evolution_equation_trace}
&\dfrac{^{\mbox{\tiny{(3)}}}R}{1-\beta} - \dfrac{2}{N}\left(1-3\frac{1+\lambda}{1-\beta}\right)D_t K + 3K^{kl}K_{kl}\notag\\
& + \dfrac{1+2\beta+3\lambda}{1-\beta}K^2 - \dfrac{4}{N(1-\beta)}D_k D^k N + \dfrac{3}{1-\beta}f(a)\notag\\
& -  \dfrac{2}{1-\beta}\chi a^2 = -\dfrac{16\pi G}{(1-\beta)c^4}\left(\mathcal{T} + \frac{2}{N}N_k \mathcal{J}^k\right)\,,
\end{align}
which will come in handy later.
		
Several comments are in order about these field equations. First, for $\lambda=\beta=0$ they reduce to those presented in Ref.~\cite{Blanchet:2011wv}\footnote{Note that
our definition of $f(a)$ as given in the action [Eqs.~\eqref{action-full-our-theory-covariant} or
\eqref{action-full-our-theory-adapted}] differs by a factor $-2$ from the definition chosen in Ref.~\cite{Blanchet:2011wv}.}. Also, the structure of these equations is clearly similar
to GR, i.e. Eq.~\eqref{eq:evolution_equation} is a modified evolution equation and Eq.~\eqref{eq:momentum_constraint} is the modified momentum constraint. On the other hand, Eq.~\eqref{eq:hamiltonian_constraint}
clearly looks like a modified Hamiltonian constraint, but a key difference from GR is present. Indeed, in GR one may in principle choose a specific gauge (defined by some conditions on $N$ and $N_i$), choose initial data
compatible with the constraints, and evolve
the evolution equation, which would ensure that the constraints are satisfied at later times. This is not possible in the case of Eqs.~\eqref{eq:hamiltonian_constraint}--\eqref{eq:evolution_equation}, since
we have already used up our ``time'' gauge freedom by adapting our coordinates to the preferred foliation. This can be seen explicitly by transforming the action of Eq.~\eqref{action-full-our-theory-covariant} to that of 
Eq.~\eqref{action-full-our-theory-adapted}, by choosing a 3+1 decomposition such that the time coordinate $t$ matches the khronon scalar $T$. As result, the lapse $N$ is \textit{not} a gauge field in Eqs.~\eqref{eq:hamiltonian_constraint}--\eqref{eq:evolution_equation},
but should rather be solved for at each step of the evolution via Eq.~\eqref{eq:hamiltonian_constraint}.
Indeed, it can be shown that once Eqs.~\eqref{eq:momentum_constraint},  \eqref{eq:evolution_equation} 
and the equations of motion of matter 
are assumed to hold, Eq.~\eqref{eq:hamiltonian_constraint} is needed to ensure the validity of
 the khronon evolution equation (which is obtained by varying the covariant action \eqref{action-full-our-theory-covariant} with respect to $T$)~\cite{Blanchet:2011wv,Jacobson:2010mx}. 
Also, as we will see below, the lack of freedom to ``gauge away'' the lapse will be the origin of the PN strong-coupling problem mentioned above.\footnote{Although in the next sections we will show that the PN dynamics becomes
non-perturbative (i.e. strongly coupled) when $\beta$ and $\lambda$ are equal or close to zero, this does
\textit{not} necessarily mean that the theory's structure \textit{itself} is pathological, even when $\beta=\lambda=0$. Indeed,
one may in principle integrate in time the evolution equation \eqref{eq:evolution_equation} as in GR, and solve the
Hamiltonian constraint \eqref{eq:hamiltonian_constraint} (which is an elliptic equation for $N$) at each time-step (given appropriate boundary conditions).
This would be possible even for $\beta=\lambda=0$, although the resulting dynamics would not be perturbatively close
to the Newtonian one.}
%

\subsection{Post-Newtonian expansion}
\label{subsec:Post-Newtonian expansion}

To calculate the  PN, let us start by writing the most generic perturbed flat metric in Cartesian coordinates $(x^0=ct,x^i)$ (see e.g. Refs.~\cite{Flanagan:2005yc,Barausse:2013ysa,Bertschinger:1993xt}):
\begin{align}\label{eq:perturbed flat metric}
g_{00} &= -1 -\frac{2}{c^2}\phi - \frac{2}{c^4}\phi_{\mbox{\tiny{(2)}}} + O(6)\notag\\
g_{0i} &= \dfrac{w_i}{c^3} + \dfrac{\partial_i \omega}{c^3} + O(5)\notag\\
g_{ij} &= \left(1 - \frac{2}{c^2}\psi\right)\delta_{ij} + \left(\partial_i \partial_j - \frac{1}{3}\delta_{ij}\nabla^2\right)\dfrac{\zeta}{c^2}\notag\\ 
&+ \dfrac{1}{c^2}\partial_{(i}\zeta_{j)} + \dfrac{\zeta_{ij}}{c^2} + O(4)\,.
\end{align} 
Under transformations of the spatial coordinates, $\psi, \zeta, \omega, \phi, \phi_{\mbox{\tiny{(2)}}}$ transform as scalars, $w_i, \zeta_{i}$ behave instead as transverse vectors (i.e. $\partial_i w^i = \partial_i\zeta^{i} = 0$),
and $\zeta_{ij}$ is a transverse and traceless tensor (i.e. $\partial_i\zeta^{ij}=\zeta_{\ i}^i=0$). 

Since we have already chosen our  time coordinate to coincide with the khronon field $T$, we only have freedom to redefine the spatial coordinates on our foliation, i.e. we are only allowed to perform gauge
transformations $h_{\mu\nu}\to h_{\mu\nu}+\partial_{(\mu} \xi_{\nu)}$, with $h_{\mu\nu}=g_{\mu\nu}-\eta_{\mu\nu}$ representing the perturbation and $\xi_{\nu}=(0,\xi_i)$ a \textit{purely spatial} vector.
For this calculation, we find it convenient to impose the gauge conditions $\zeta = \zeta_i = 0$~\cite{Barausse:2013ysa}. As a result, the lapse, shift, spatial metric and acceleration at 1PN order
are given by 
\begin{gather}
N = \dfrac{1}{\sqrt{-g^{00}}}= 1 + \dfrac{\phi}{c^2} - \dfrac{1}{2}\dfrac{\phi^2}{c^4} + \dfrac{\phi_{\mbox{\tiny{(2)}}}}{c^4} +O(6), \\ 
N_i = g_{0i}=\dfrac{w_i}{c^3} + \dfrac{\partial_i \omega}{c^3} + O(5)\,, \\
\gamma_{ij} = g_{ij} = \left(1 - \frac{2}{c^2}\psi\right)\delta_{ij} + \dfrac{\zeta_{ij}}{c^2} + O(4)\,,\\
a_i=\dfrac{\partial_i\phi}{c^2} -2\dfrac{\phi\partial_i\phi}{c^4} + \dfrac{\partial_i\phi_{\mbox{\tiny{(2)}}}}{c^4}+ O(6)\label{acceleration}
\end{gather}
which can be used to compute the left-hand sides of the field equations \eqref{eq:hamiltonian_constraint}--\eqref{eq:evolution_equation_trace}. To compute the right-hand side of those equations,
we use a perfect fluid stress-energy tensor, i.e.
\begin{equation}
T^{\mu\nu} = \biggl(\rho + \dfrac{p}{c^2}\biggr)u^{\mu}u^{\nu} + pg^{\mu\nu},
\end{equation}
where $\rho$ is the matter mass-energy density, $p$ the pressure and $u^{\mu} = dx^{\mu}/d\tau$ the four-velocity of the fluid elements (with $\tau$ the proper time).

Before proceeding with the calculation, let us clarify the PN order of the function $f(a)$. As mentioned in section~\ref{sec:Einstein-Aether and khronometric theories}, 
in the high-acceleration regime (i.e. for $a c^2\gg a_0$),
$f(a)\approx  -2\Lambda+\alpha a^2\approx \alpha a^2$, so Eq.~\eqref{acceleration} implies $f(a)=O(4)$. Note that in deriving this scaling
we have used the fact that $\Lambda$ is comparable to the observed value of
the cosmological constant, i.e. $c^4\Lambda\sim c^2H_0^2\sim a_0^2\ll a^2c^4$, which allows neglecting the $-2\Lambda$ term. (Of course, this corresponds to the known
fact that the  cosmological constant has negligible impact on the 1PN dynamics on small scales.)
In the low-acceleration regime (i.e. on cosmological scales), the cosmological constant would instead be expected to enter the 1PN dynamics. Indeed, 
for $a c^2\ll a_0$, $f(a)\approx  -2\Lambda_0+2a^2-4 a^3 c^2/(3 a_0)$, and assuming (as is natural to do) that $\Lambda_0$ is comparable to the observed
value of the cosmological constant, the term $-2\Lambda_0$ dominates over $2a^2-4 a^3 c^2/(3 a_0)=O(4)$. However, in order to have the same scaling $f(a)=O(4)$
as in the high-acceleration regime, we can simply move the cosmological constant to the right-hand side of the field equations, and absorb it in the matter stress-energy tensor
as a ``fluid'' component with equation of state $p/c^2=-\rho=-\Lambda c^2/(8\pi G)$, as routinely done in cosmology. Therefore, in what follows we will consider $f(a)=O(4)$ in
both the high- and low-acceleration regimes, with the caveat that in the latter $f(a)\approx  2a^2-4 a^3 c^2/(3 a_0)$ \textit{and} the matter is meant to include a
``Dark-Energy'' component $p/c^2=-\rho=-\Lambda c^2/(8\pi G)$.

With these Ans\"atze and scalings, deriving the 1PN field equations is now straightforward. In particular, expanding 
Eq.~\eqref{eq:evolution_equation_trace}  to lowest order in $1/c$ yields~\cite{Blanchet:2011wv}
\begin{equation}\label{eq:quasi_equality}
\psi = \phi + O(2)\,,
\end{equation}
which implies that light deflection behaves as in GR (except, as we will show below, that the relation between $\phi$ and the mass distribution of matter is different than in GR). 
This is important as it allows the theory to reproduce the successes of the $\Lambda$CDM model in the interpretation of gravitational lensing 
from galaxies and clusters of 
galaxies~\cite{Blanchet:2011wv,Bekenstein:1993fs,Famaey:2011kh}. Also, based on Eq.~\eqref{eq:quasi_equality},
we can write 
\begin{equation}
\psi = \phi +\frac{\delta\psi}{c^2}+ O(4)\,,
\end{equation}
where we have defined the potential $\delta\psi$, which will appear in the rest of the calculation [c.f. Eq.~\eqref{eq:delta_psi} below].

Using this result in Eq.~\eqref{eq:hamiltonian_constraint}, to lowest order in $1/c$ we obtain~\cite{Blanchet:2011wv}
\begin{equation}\label{eq:modified poisson equation with chi}
\vec{\nabla} \cdot \left[\left(1-\dfrac{\chi}{2}\right)\vec{\nabla}\phi\right] = 4\pi G \rho + O(2)\,,
\end{equation}
where since $f(a)\propto a^2=O(4)$, one has that $\chi=f'(a)/(2 a)$ is of zeroth-order in $1/c$.
In the high-acceleration regime, $f(a)\approx \alpha a^2$, thus $\chi=\alpha$ and this equation
becomes the usual Poisson equation
\begin{equation}\label{eq:poisson}
{\nabla}^2\phi_N = 4\pi G_N \rho + O(2)\,,
\end{equation}
with $G_N$ given by Eq.~\eqref{GN}.
At intermediate and low accelerations, $\chi$ is not necessarily constant, and defining
an ``interpolation function''
\begin{equation}
\mu = 1-\dfrac{\chi}{2}\,,
\end{equation}
Eq.~\eqref{eq:modified poisson equation with chi} becomes the modified Poisson equation of the MOND dynamics~\cite{Milgrom:1983ca,Milgrom:1983pn,Milgrom:1983zz,Famaey:2011kh}, i.e.
\begin{equation}\label{eq:modified_poisson}
\vec{\nabla} \cdot \biggl[\ \mu\left(\dfrac{|\vec{\nabla}\phi|}{a_0}\right) \; \vec{\nabla}\phi \ \biggr] = 4\pi G \rho + O(2).
\end{equation}
In particular, in the low-acceleration regime $ac^2\ll a_0$ (i.e. in the ``deep-MOND regime''), $f(a)\approx 2a^2-4 a^3 c^2/(3 a_0)$ and Eq.~\eqref{eq:modified poisson equation with chi} becomes
\begin{equation}\label{eq:deepMOND}
\vec{\nabla} \cdot \left(\dfrac{|\vec{\nabla}\phi|}{a_0}\vec{\nabla}\phi\right) = 4\pi G\rho\,.
\end{equation}

From the off-diagonal part of the modified evolution equation [Eq.~\eqref{eq:evolution_equation}] we obtain
\begin{equation}
\bm{\zeta} _{ij} = O(2)\,,
\end{equation}
i.e. $\bm{\zeta} _{ij}$ appears at higher order than 1PN. This is of course expected, since this term represents gravitational waves,
which do not enter in the 1PN metric in GR.

Solving then the trace of the evolution equation [Eq.~\eqref{eq:evolution_equation_trace}] to $O(4)$, we obtain
\begin{align}\label{eq:delta_psi}
&\dfrac{2}{c^4}\nabla^2 \delta\psi = -\dfrac{3}{2}f(a) + \dfrac{1}{c^4}\biggl(-24\pi G p - 8\pi\rho v^2 - 7\partial_i \phi \partial_i \phi \notag\\
&+\chi\partial_i \phi \partial_i \phi - 8\phi\nabla^2\phi + (2 + \beta + 3\lambda)(\partial_t\nabla^2\omega + 3\partial_t^2 \phi)\biggr)\,,
\end{align}
and by replacing this expression in the modified Hamiltonian constraint [Eq.~\eqref{eq:hamiltonian_constraint}] at $O(4)$
we find
\begin{align}\label{eq:modified_poisson_1PN}
&\vec{\nabla} \cdot \Biggl[\biggl(1-\dfrac{\chi}{2}\biggr)\vec{\nabla} \biggl(\phi + \dfrac{\phi_{\mbox{\tiny{(2)}}}}{c^2}\biggr)\Biggr] = 4\pi G \rho + c^2\dfrac{f(a)}{2}\notag\\
& + \dfrac{1}{c^2}\Bigl(8\pi G \rho v^2 + 12\pi G p + 2\vec{\nabla}\phi\cdot\vec{\nabla} \phi - \dfrac{3}{2}\chi\vec{\nabla}\phi\cdot\vec{\nabla} \phi\notag\\
& - \dfrac{1}{2}(2+\beta+3\lambda)\bigl(\partial_t\nabla^2 \omega + 3\partial_t^2\phi\bigr)\Bigr).
\end{align}
Finally, the 1PN equation for the ``frame-dragging'' potential $w_i$ can be obtained from the momentum constraint [Eq.~\eqref{eq:momentum_constraint}], whose 
expansion yields 
\begin{align}\label{eq:momentum constraint 1PN}
\nabla^2 w_i + 2\Bigl(\dfrac{\beta+\lambda}{\beta -1}\Bigr)&\partial_i \nabla^2 \omega\notag\\
& = \dfrac{16\pi G \rho v_i }{1-\beta} -2\Bigl(\dfrac{2+\beta+ 3\lambda}{\beta -1}\Bigr)\partial_i \partial_t\phi\,.
\end{align}
By taking the divergence of this equation we obtain
\begin{equation}\label{eq:momentum constraint 1PN_SC}
\nabla^2 \nabla^2 \omega = \frac{1}{\beta+\lambda}[{8\pi G \partial_t \rho} -({2+\beta+ 3\lambda})\partial_t\nabla^2\phi]\,,
\end{equation}
where we have used the condition  $\partial_i w^i=0$ (c.f. the definition of $w_i$), 
and the energy conservation to Newtonian order, $\partial_t\rho=-\partial_i(\rho v^i) [1+O(2)]$.
Denoting by $\phi_N=4\pi G_N \nabla^{-2}\rho$ [with $G_N$ given by Eq.~\eqref{GN}] the Newtonian potential in the high-acceleration regime, we can rewrite
Eq.~\eqref{eq:momentum constraint 1PN_SC}
in the more useful form
\begin{equation}\label{eq:momentum constraint 1PN_SC_bis}
\nabla^2 \omega = \frac{1}{\beta+\lambda}\partial_t[(2-\alpha) \phi_N -({2+\beta+ 3\lambda})\phi]\,.
\end{equation}
Note that Eqs.~\eqref{eq:modified_poisson_1PN}--\eqref{eq:momentum constraint 1PN_SC_bis} 
are valid both in the high-acceleration regime, in which $f(a)\approx \alpha a^2$, and in the low-acceleration, deep-MOND regime, characterized
by $f(a)\approx 2a^2-{4 a^3 c^2}/(3 a_0)$ (and thus $\chi=2 - {2c^2 a}/{a_0}$). 
In the high-acceleration regime,
we must of course recover the known results for khronometric theory, namely that
all the PPN parameters vanish except for $\alpha_1$ and $\alpha_2$, which are given by Eqs.~\eqref{eq:alpha1} and \eqref{eq:alpha2}.
Indeed, we show explicitly that this is the case in the Appendix.

The low-acceleration, deep-MOND regime is instead analyzed in detail in the next section. 
However, already looking at Eq.~\eqref{eq:momentum constraint 1PN_SC_bis},
we can understand that the 1PN expansion in the deep-MOND regime may
have a non-perturbative character, because the right-hand side seems to diverge for $\beta+\lambda\to0$.
Clearly, this cannot be the case in GR, where we know that the 1PN expansion is perturbative. Indeed, in GR one has
 $\alpha=\beta=\lambda=0$ and $\phi=\phi_N$, thus the two terms in round brackets on the right-hand side cancel out. This
is consistent with the fact that in GR one can set $\omega=0$ by a gauge transformation of the time coordinate~\cite{Barausse:2013ysa} (while still imposing
the conditions $\zeta = \zeta_i = 0$ by a gauge transformation of the spatial coordinates, as we do in this paper).
Because in the khronometric theories that we are considering we already fixed the time foliation by adapting it to the khronon $T$, we have no residual
gauge freedom to set $\omega$ to zero, and $\nabla^2 \omega$ may indeed diverge (in general) when $\lambda,\beta\to 0$.
Another way of seeing that the case $\beta=\lambda=0$ is pathological is to note that if we had started from such a theory, we would
have derived Eq.~\eqref{eq:momentum constraint 1PN} with $\beta=\lambda=0$, i.e. the same equation as in GR. That equation, however,
would have no dependence on $\omega$, which would therefore remain completely undetermined. This is not a problem in GR, as $\omega$ is a gauge mode (so it should
indeed remain undetermined), but \textit{is} a problem in the modified khronometric theory of Ref.~\cite{Blanchet:2011wv}, because $\omega$ is \textit{not} a gauge mode there.

Indeed, already in the high-acceleration regime the terms in round
brackets on the right-hand side of Eq.~\eqref{eq:momentum constraint 1PN_SC_bis} do \textit{not} cancel out (in general)
in the theories we are considering.
This is because $\phi=\phi_N$ is that regime, but $\alpha, \beta$ and $\lambda$ are in general non-zero. Of course 
this corresponds to the fact that in khronometric theory the preferred-frame parameter $\alpha_2$ becomes large when $\lambda+\beta$
is small, unless $\alpha\approx 2\beta$ [c.f. Eq.~\eqref{eq:alpha2}].\footnote{Indeed, for $\alpha= 2 \beta$ and $\phi=\phi_N$, the right-hand side of Eq.~\eqref{eq:momentum constraint 1PN_SC_bis}
is independent of $\beta,\lambda$.}
 For high accelerations, however, we have already discussed 
 that one does indeed have the freedom to set $\alpha\approx 2\beta$, so as to satisfy the solar-system constraints
$\vert\alpha_1\vert \lesssim 10^{-4}$ and $\vert\alpha_2\vert \lesssim 10^{-7}$. There is therefore no strong-coupling 
problem in the viable part of the parameter space of the couplings at high accelerations.

The situation is different in the low-acceleration, deep-MOND regime, since $\phi\neq\phi_N$ there. 
Indeed, we will show explicitly that the right-hand side of Eq.~\eqref{eq:momentum constraint 1PN_SC_bis}
diverges in the limit $\beta,\lambda\to 0$,  in low-acceleration, time-dependent/dynamical systems. We will also show that this strong-coupling problem
appears in a region of parameter space that would be otherwise allowed based on experimental constraints
coming from the high-acceleration regime.

\subsection{The strong-coupling problem in galactic rotation curves}
\label{sec:Simplified physical system}

As shown in the end of the previous section, the 1PN equations present a strong-coupling problem in time-dependent situations, if $\lambda$ and $\beta$ are very close
to $0$. In this section we will show this explicitly by solving the 1PN equations for a toy model consisting of
 a spherical galaxy whose mass $M(t)$ increases linearly as a function of time due to e.g. accretion of gas from the IGM. 
 We will then compute the conditions that $\lambda$ and $\beta$ must satisfy
to avoid this strong-coupling problem, and show the resulting parameter space in which the theory remains viable.
More specifically, we will compute the rotation curves for such an accreting galaxy outside its luminous center, assuming that no
Dark Matter is present, and assess for what values of $\lambda$ and $\beta$ the aforementioned strong-coupling problem 
modifies the rotation curves in a way that is incompatible with observations~\cite{Rubin:1980zd,Sofue:2000jx,1991MNRAS.249..523B,Persic:1995ru}.

\subsubsection{The Newtonian order}
\label{subsec:0PN equation}

At Newtonian order, the equation for the perturbation $\phi$ at low accelerations $|\vec{\nabla} \phi|\ll a_0 $
is given by Eq.~\eqref{eq:deepMOND}. In spherical symmetry, however, $\phi$ is only a function of the distance $r$ from the galaxy's center, thus
we can always write $|\vec{\nabla} \phi| {\vec{\nabla} \phi}/{a_0}=\vec{\nabla} S$ for some scalar function $S(r)$. Inserting this definition in
Eq.~\eqref{eq:deepMOND}, we obtain that $S$ must coincide with the  GR Newtonian potential $\phi_N$. Therefore, to find the MOND gravitational potential 
$\phi$ in spherical symmetry, we can simply solve the corresponding Newtonian problem in GR for $\phi_N$, and then
compute $\phi$ by solving
\begin{equation}\label{eqPhi}
\frac{\mbox{d}\phi(r)}{\mbox{d}r}=\sqrt{a_0 \frac{\mbox{d}\phi_N(r)}{\mbox{d}r}}\,.
\end{equation}

As our toy model for an accreting galaxy, let us consider a spherical body with mass $M=M_0+\dot{M} t$ (with $\dot{M}$ and $M_0$ constants)
and radius $R$,  
surrounded by a spherically symmetric, stationary accretion flow [whose density, simply by mass conservation, is $\rho= \dot{M}/[4 \pi r^2 v_r(r)]$,
where $v_r(r)$ is the radial infall velocity as a function of radius]. 
Note that because no Dark Matter is assumed to exist, we identify $R$ with the galaxy's half-mass radius, which is related to
the (baryonic) mass $M$ by the observational fit~\cite{shenfit}
\begin{equation}\label{eq:fit}
\log_{10}(R_{\rm eff}/\mbox{kpc})=
\begin{cases}
-5.54 + 0.56\log_{10}\left(\frac{M}{M_\odot}\right) \\
\qquad \ \text{for} \log_{10}\left(\frac{M}{M_\odot}\right) > 10.3,\\
\ \\
-1.21 + 0.14 \log_{10}\left(\frac{M}{M_\odot}\right)  \\
 \qquad \  \text{for} \log_{10}\left(\frac{M}{M_\odot}\right) \leq 10.3 .
\end{cases}
\end{equation}
Let us focus on the region outside the galaxy's radius $R$, where
only the accreting gas and the cosmological constant are present. 
In this region, $\phi_N$ is given by
\begin{equation}\label{eq:PhiN}
\phi_N= -\frac{G_N M}{r}+{\cal O}_{\rm finite}(\dot{M}, \Lambda_{\rm obs})\,,
\end{equation}
where ${\cal O}_{\rm finite}(\dot{M}, \Lambda_{\rm obs})$ denotes corrections (proportional to either $\Lambda_{\rm obs}$ or 
$\dot M$) that remain \textit{finite} as $\beta, \lambda\to 0$. Indeed, these corrections are clearly independent of $\beta, \lambda$ in this case,
and are also time-independent,
because both $\rho=\Lambda_{\rm obs}c^2/(8 \pi G_N)$ and $\rho= \dot{M}/[4 \pi r^2 v_r(r)]$ do not change with time.

To compute $\phi$, one can then just solve Eq.~\eqref{eqPhi}. To do so, one needs to specify conditions
ensuring a smooth transition to the GR solution, which is valid in the high-acceleration regime near the galaxy. In 
particular, let us define the transition radius
\begin{equation}
 r_0 = \sqrt{\dfrac{G_N M}{a_0}}
\end{equation}
at which the Newtonian gravitational acceleration $|\vec{\nabla} \phi_N| = GM/r^2+{\cal O}_{\rm finite}(\dot{M}, \Lambda_{\rm obs})$ 
matches the acceleration constant $a_0$. [Note that $r_0$ is larger than the half-light radius given  by Eq.~\eqref{eq:fit} for typical galaxy masses.]

At distances from the body's center  $r\ll r_0$ (but $r>R$, i.e. outside the galaxy), $\phi$ coincides with $\phi_N$ as given
by Eq.~\eqref{eq:PhiN}, while for $r\gg r_0$, $\phi$ is given by Eq.~\eqref{eqPhi}. 
We can therefore assume a sharp transition at $r=r_0$, and solve Eq.~\eqref{eqPhi} by imposing 
continuity of $\phi$ and its first derivative, i.e. $\phi (r_0) = - G_NM/r_0+{\cal O}_{\rm finite}(\dot{M}, \Lambda_{\rm obs})$ and $d\phi/dr = G_N M/r_0^2+{\cal O}_{\rm finite}(\dot{M}, \Lambda_{\rm obs})$, thus obtaining
\begin{equation}\label{eq:mond_potential}
\phi = \sqrt{G_N M a_0} \left( \ln \left(\dfrac{r}{r_0}\right) - 1\right)+{\cal O}_{\rm finite}(\dot{M}, \Lambda_{\rm obs})
\end{equation}
for $r>r_0$.

Finally, as we will show explicitly in the next section, we do \textit{not}
need the explicit form of the potential $\phi=\phi_N$
inside the galaxy (i.e. for $r<R\ll r_0$) to solve the 1PN equations,
if we focus on the terms that dominate when $\beta+\lambda\to 0$.

\subsubsection{The metric at 1PN order}
\label{subsec:1PN equations}

At 1PN order, the metric is characterized by the potentials $\omega$, $w^i$ and $\phi_{\mbox{\tiny{(2)}}}$. 
In spherical symmetry, however, $w^i=0$.\footnote{This follows from the requirement that
$\partial_i w^i=0$, imposing regularity at $r=0$. Alternatively, one can solve 
the divergenceless part of Eq.~\eqref{eq:momentum constraint 1PN}, noting that 
in spherical symmetry the velocity only has a radial component $v_r(r)$, which
can be expressed as the gradient of a scalar potential.}

To determine $\omega$, let us start from Eq.~\eqref{eq:momentum constraint 1PN_SC_bis}. By using the Green function of the Laplace operator we obtain
\begin{multline}\label{eq:omega_cartesian}
\omega(\vec{x},t) =\\ -\partial_t\left[\int_{r'>r_0} d^3 \vec{x}' \dfrac{(2-\alpha)\phi_N(r',t) - (2+\beta+3\lambda)\phi(r',t)}{4\pi(\beta+\lambda)|\vec{x} - \vec{x}'|}\right.\\\left.
-\int_{r'<r_0} d^3 \vec{x}' \dfrac{(\alpha+\beta+3\lambda)\phi_N(r',t)}{4\pi(\beta+\lambda)|\vec{x} - \vec{x}'|}\right]+\psi_0\,,
\end{multline}
where $\psi_0$ is an integration constant, and we have used the fact that $\phi=\phi_N$ at high accelerations (i.e. for $r<r_0$). 
As already noted in the previous section, if $\alpha\approx 2\beta$ (as required by solar-system tests), the second integral 
in Eq.~\eqref{eq:omega_cartesian} is finite when $\beta,\lambda\to 0$, and can therefore be neglected with respect to the first one, which diverges. 
More precisely, by assuming $\alpha= 2\beta+{\cal O}(\alpha_1,\alpha_2)$ (so as to pass solar-system tests), in spherical coordinates
the above solution becomes
\begin{multline}\label{eq:omegaa}
\omega(r,t) =  -\dfrac{1}{(\beta+\lambda)}\times \\
\partial_t\bigg\{\dfrac{1}{r}\int_{r_0}^r dr' r'^2[2(1-\beta)\phi_N(r',t) - (2+\beta+3\lambda)\phi(r',t)]\\
 + \int_r^{R_\textup{max}} dr' r' [2(1-\beta)\phi_N(r',t) - (2+\beta+3\lambda)\phi(r',t)]\bigg\}\times\\\times \left[1+{\cal O}(\alpha_1,\alpha_2)\right]
+\psi_0+{\cal O}(\beta+\lambda)^0\,.
\end{multline}
Because $\phi$ diverges as $\ln r$ as $r\to\infty$,
the second integral on the right-hand side of this equation formally diverges. This is simply because
the PN formalism is by definition a perturbative expansion on a Minkowski background [c.f. Eq.~\eqref{eq:perturbed flat metric}]. 
Of course, for any given spacetime one can choose locally Riemannian coordinates $x^\alpha$ centered on a given event, and such
that the metric is locally $g_{\mu\nu}=\eta_{\mu\nu}+{\cal O}(r/{\cal R})^2$, where $r\approx\sqrt{\eta_{\alpha\beta} x^\alpha x^\beta}$ 
is the proper distance from the 
event and ${\cal R}$ is the curvature radius of the spacetime at the event. In the particular case of a system embedded in a 
cosmological spacetime, ${\cal R}\sim c/H$ ($H$ being the Hubble rate), i.e. 
the Minkowski metric is the appropriate background metric only on length- and time-scales much smaller than the
cosmological ones (i.e., respectively, the Hubble radius and Hubble time)~\cite{Will:1993ns}. 
For this reason, we can truncate the second integral on the right-hand side of Eq.~\eqref{eq:omegaa} at a cut-off radius $R_{\max}$,
which can be thought of as much smaller than the 
present Hubble radius but much larger than the typical size of the luminous component of a galaxy.

In practice, the cut-off $R_{\max}$ never enters our calculations and results, as it can be renormalized in the integration
constant $\psi_0$. Indeed, once this cut-off is imposed, we 
can use Eqs.~\eqref{eq:PhiN} and~\eqref{eq:mond_potential} for $\phi_N$ and $\phi$ in Eq.~\eqref{eq:omegaa}, and the integration yields the following expression
\begin{align}
&\omega(r,t) = -\dfrac{\dot{M}}{72 r (\beta+\lambda)}\biggl[72G_N(1-\beta)\left(r^2+r_0^2\right)\notag\\
& - (2+\beta+3\lambda)\sqrt{\dfrac{a_0 G_N}{M(t)}}\left(17r^3 +28r_0^3- 6r^3 \ln\left(\dfrac{r}{r_0}\right)\right)\biggr]\notag\\&
-\dfrac{\dot{M} R_{\max}}{8(\beta+\lambda)}\biggl[-16 G_N(1-\beta)+ R_{\max} (2+\beta+3\lambda)\times\notag\\&\times\sqrt{\frac{a_0 G_N}{M}} (5+2 \ln(r_0/R_{\max}) \biggr]
\times\left[1+{\cal O}(\alpha_1,\alpha_2)\right]
+\psi_0\notag\\&+{\cal O}_{\rm finite}(\dot{M}, \Lambda_{\rm obs})+{\cal O}(\beta+\lambda)^0\,,
\end{align}
from which it is clear that the terms that depend on the cut-off radius can be absorbed in the integration constant $\psi_0$. Therefore, the final solution
for the potential  $\omega(r,t)$ is simply
\begin{align}
&\omega(r,t) = -\dfrac{\dot{M}}{72 r (\beta+\lambda)}\biggl[72G_N(1-\beta)\left(r^2+r_0^2\right)\notag\\
& - (2+\beta+3\lambda)\sqrt{\dfrac{a_0 G_N}{M(t)}}\left(17r^3 +28r_0^3- 6r^3 \ln\left(\dfrac{r}{r_0}\right)\right)\biggr]\notag\times\\&
\times \left[1+{\cal O}(\alpha_1,\alpha_2)\right]+{\cal O}_{\rm finite}(\dot{M}, \Lambda_{\rm obs})+{\cal O}(\beta+\lambda)^0\,,\label{eq:w}
\end{align}

Let us now consider the modified Hamiltonian constraint Eq.~\eqref{eq:modified_poisson_1PN}. Because of spherical symmetry and taking into account only the terms that diverge when $\beta+\lambda \rightarrow 0$, at 1PN order that equation becomes
\begin{align}\label{eq:1PN hamiltonian in progress}
&\dfrac{2}{a_0 r^2}\Biggl[\dfrac{\partial}{\partial r}\Biggl(r^2 \dfrac{\partial \phi}{\partial r} \dfrac{\partial \phi_{\mbox{\tiny{(2)}}}}{\partial r}\Biggr)\Biggr]\notag \\&=- \dfrac{2+\beta+3\lambda}{2} \partial_t \nabla^2 \omega +{\cal O}(\beta+\lambda)^0= \notag \\&
- \dfrac{2+\beta+3\lambda}{2(\beta + \lambda)}\partial^2_t \left[2(1-\beta)\phi_N -(2+\beta+3\lambda)\phi\right]\notag \times\\&\times \left[1+{\cal O}(\alpha_1,\alpha_2)\right]+{\cal O}(\beta+\lambda)^0
\end{align}
By inserting the explicit expression for $\phi$ at low accelerations [Eq.~\eqref{eq:mond_potential}] and 
isolating the derivatives of $\phi_{\mbox{\tiny{(2)}}}$ on the left-hand side, in the deep-MOND region $r>r_0$, Eq.~\eqref{eq:1PN hamiltonian in progress}
becomes
\begin{align}\label{eq:diff_eq}&
\dfrac{\partial}{\partial r}\Biggl(r \dfrac{\partial \phi_{\mbox{\tiny{(2)}}}}{\partial r}\Biggr)=F(r,t)\equiv\notag\\&
- \dfrac{(2+\beta+3\lambda) r^2}{4(\beta + \lambda) r_0}\partial^2_t \left[2(1-\beta)\phi_N -(2+\beta+3\lambda)\phi\right]\notag\times \\&\times \left[1+{\cal O}(\alpha_1,\alpha_2)\right]+{\cal O}(\beta+\lambda)^0+{\cal O}_{\rm finite}(\dot{M}, \Lambda_{\rm obs})
\end{align}
where $F(r, t)$ represents the source on the right-hand side. 
To solve this equation, let us construct the Green function ${G}(r,r')$,
i.e. the solution to ${\partial_ r}(r \partial_r G)=\delta(r-r')$. 
As usual, the Green function can be constructed from solutions of the homogeneous problem. In brief, for $r\neq r'$, the equation
defining the Green function becomes ${\partial_ r}(r \partial_r G)=0$, which
has the general solution ${G}(r,r')=K_2 \ln\left({r}/{r_0}\right)  + K_1$, with $K_1$ and $K_2$
being integration constants. Imposing then the junction conditions $G|_{r=r'+0^+}=G|_{r=r'-0^+}$ and $r \partial_r G|_{r=r'+0^+}-r \partial_r G|_{r=r'+0^-}=1$ to account for the presence of the Dirac delta on the right-hand side, we then obtain
\begin{equation}
{G}(r,r')=
\begin{cases}
K_2 \ln\left(\dfrac{r}{r_0}\right)  + K_1  & r < r'\\
\ln\left(\dfrac{r}{r'}\right) + K_2 \ln\left(\dfrac{r}{r_0}\right) + K_1  & r > r'
\end{cases}
\end{equation}
The general solution to Eq.~\eqref{eq:diff_eq} can then be written as
\begin{align}
\phi_{\mbox{\tiny{(2)}}} (r, t) &=  \left[K_2 \ln\left(\dfrac{r}{r_0}\right)  + K_1\right]\int_{r_0}^{R_\textup{max}} dr' F(r', t)\notag\\
& + \int_{r_0}^{r} dr' \ln\left(\dfrac{r}{r'}\right) F(r', t)
\end{align}
which explicitly gives 
\begin{align}
&\phi_{\mbox{\tiny{(2)}}} (r, t) = -\dfrac{a_0(2+\beta+3\lambda)^2\dot{M}^2}{432 (\beta+\lambda)M^2}\biggl[5(r_0^3-r^3)\notag\\
&+12K_1(r_0^3-R_\textup{max}^3)+9K_1 R_\textup{max}^3\ln\left(\dfrac{R_\textup{max}}{r_0}\right)\notag\\
& + 3\ln\left(\dfrac{r}{r_0}\right)\biggl(r^3+4r_0^3 + 4K_2(r_0^3-R_\textup{max}^3)\notag\\
&+3K_2R_\textup{max}^3\ln\left(\dfrac{R_\textup{max}}{r_0}\right)\biggr)\biggr]\times \left[1+{\cal O}(\alpha_1,\alpha_2)\right]\notag\\&
+{\cal O}(\beta+\lambda)^0+{\cal O}_{\rm finite}(\dot{M}, \Lambda_{\rm obs})\,.
\end{align}
The integration constants $K_1$ and $K_2$ can be fixed by imposing that the solution matches the high-acceleration solution $\phi^{\rm high \ acc}_{\mbox{\tiny{(2)}}}$
and its derivative
at $r=r_0$, i.e. $\phi_{\mbox{\tiny{(2)}}} (r_0) = \phi^{\rm high \ acc}_{\mbox{\tiny{(2)}}}(r_0)\equiv H_1$ and 
$\partial_r \phi_{\mbox{\tiny{(2)}}}(r_0) =\partial_r \phi^{\rm high \ acc}_{\mbox{\tiny{(2)}}}(r_0)\equiv H_2$, thus obtaining
\begin{align}
&\phi_{\mbox{\tiny{(2)}}} (r, t) = H_1 + H_2 r_0\ln\left(\dfrac{r}{r_0}\right) - \dfrac{a_0(2+\beta+3\lambda)^2\dot{M}^2}{432(\beta+\lambda)M^2}\times\notag\\
&\times\biggl(5(r_0^3-r^3)+3(r^3+4r_0^3)\ln\left(\dfrac{r}{r_0}\right)\biggr)\notag\times\\&\times\left[1+{\cal O}(\alpha_1,\alpha_2)\right]+{\cal O}(\beta+\lambda)^0+{\cal O}_{\rm finite}(\dot{M}, \Lambda_{\rm obs})\,.\label{eq:phi2}
\end{align}
Note that the cut-off radius $R_{\max}$ is once again absorbed in the integrations constants $H_1$ and $H_2$ (as for the potential $\omega$ earlier in this section).
Since in the high-acceleration regime the theories that we consider reduce to khronometric theory, in which no strong-coupling problem 
is present when $\beta+\lambda\to 0$ if the solar-system constraints are satisfied (c.f. discussion in section \ref{subsec:Post-Newtonian expansion}),
we have $H_1={\cal O}(\beta+\lambda)^0$ and $H_2={\cal O}(\beta+\lambda)^0$, and Eq.~\eqref{eq:phi2} can be rewritten simply as
\begin{align}
&\phi_{\mbox{\tiny{(2)}}} (r, t) = - \dfrac{a_0(2+\beta+3\lambda)^2\dot{M}^2}{432(\beta+\lambda)M^2}\times\notag\\
&\times\biggl(5(r_0^3-r^3)+3(r^3+4r_0^3)\ln\left(\dfrac{r}{r_0}\right)\biggr)\times\notag\\&\times\left[1+{\cal O}(\alpha_1,\alpha_2)\right] +{\cal O}(\beta+\lambda)^0+{\cal O}_{\rm finite}(\dot{M}, \Lambda_{\rm obs})\,.\label{eq:phi2final}
\end{align}

\subsubsection{The impact of the strong coupling on the rotation curves}
\label{subsec:Test particle on circular motion}

Let us now explore the impact of the strong-coupling problem described above on the rotation curves of galaxies. To this
purpose, let us model the gas (whose velocity is measured to determine the rotation curves) by test particles following circular geodesics
in the deep-MOND region $r>r_0$. Because of spherical symmetry, we can assume that the orbits are
on the equatorial plane, without loss of generality, i.e. in spherical coordinates $x^{\mu} = (ct, r, \theta, \varphi)$ the
four-velocity of the gas is
\begin{equation}
u^{\mu}  = \dfrac{d t}{d\tau}\left(c, 0,0,\dot{\varphi}\right),
\end{equation}
where at 1PN order the relation between coordinate time $t$ and proper time $\tau$ is given by
\begin{equation}
\dfrac{d t}{d\tau} = 1 - \dfrac{\phi}{c^2} + \dfrac{(r\dot{\varphi})^2}{2c^2} + O(4)\,,
\end{equation}
which follows from the normalization condition $u_\mu u^\mu=-c^2$.

The geodesics equation
\begin{equation}
\dfrac{d^2 x^{\mu}}{d\tau^2} = -\Gamma_{\alpha\beta}^{\mu} \dfrac{dx^\alpha}{d\tau}\dfrac{dx^{\beta}}{d\tau}
\end{equation}
can now be expressed in terms of the coordinate time (i.e. the time measured by an observer far from the galaxy):
\begin{equation}\label{eq:coordinate acceleration}
\dfrac{d^2 x^{\mu}}{dt^2} = -\Gamma_{\alpha\beta}^{\mu} \dfrac{dx^\alpha}{dt}\dfrac{dx^{\beta}}{dt} + \dfrac{1}{c} \dfrac{d x^{\mu}}{dt}\Gamma_{\alpha\beta}^t \dfrac{dx^{\alpha}}{dt}\dfrac{dx^{\beta}}{dt}\,.
\end{equation}
Focusing on the radial component, and because $d^2r/dt^2=dr/dt=0$ for circular orbits, Eq.~\eqref{eq:coordinate acceleration} then gives
\begin{align}\label{eq:squared_velocity}
&v_{\varphi, \mbox{\tiny{1PN}}}^2 = r^2 \dot{\varphi}^2 = r\dfrac{\partial \phi}{\partial r}(t,r) + \frac{r}{c^2}\biggl[r \left(\dfrac{\partial \phi}{\partial r}(t,r)\right)^2\notag\\
& + 2 \phi (t,r) \dfrac{\partial \phi}{\partial r}(t,r) + \dfrac{\partial \phi_{\mbox{\tiny{(2)}}}}{\partial r}(t,r)  + \dfrac{\partial^2 \omega}{\partial t\partial r}(t,r)\biggr]+O(4).
\end{align}

By using Eq.~\eqref{eq:mond_potential},
at the lowest (i.e. Newtonian) order, this equation yields 
\begin{equation}\label{eq:Tully-Fisher}
v^2_{\varphi,\mbox{\tiny{N}}} = \sqrt{G_N M a_0}+{\cal O}_{\rm finite}(\dot{M}, \Lambda_{\rm obs})+O(2)\,,
\end{equation}
i.e. the rotation curves of galaxies are flat in the deep-MOND region. [Note also that the scaling
of Eq.~\eqref{eq:Tully-Fisher} with the mass
agrees with the Tully-Fisher relation for disk galaxies and the Faber-Jackson relation for elliptical galaxies and clusters, c.f. 
Ref.~\cite{Famaey:2011kh} for a review of these two relations
in the context of MOND.]
At 1PN order, and focusing on the terms that diverge as $\beta+\lambda\to 0$,
the rotational velocity becomes
\begin{align}\label{eq:squared_velocity_approx}
&v_{\varphi, \mbox{\tiny{1PN}}}^2 = \sqrt{G_N M a_0} + \frac{r}{c^2}\biggl( \dfrac{\partial \phi_{\mbox{\tiny{(2)}}}}{\partial r}(t,r)  + \dfrac{\partial^2 \omega}{\partial t\partial r}(t,r)\biggr)\notag\\&+{\cal O}(\beta+\lambda)^0+{\cal O}_{\rm finite}(\dot{M}, \Lambda_{\rm obs})+O(4)
\end{align}
or more explicitly,
by using the solutions given by Eqs.~\eqref{eq:w} and \eqref{eq:phi2}, 
\begin{align}\label{eq:velocita_1PN}
&v_{\varphi, \mbox{\tiny{1PN}}}^2 = \sqrt{G_N M(t) a_0} \notag\\&+ \frac{1}{c^2}\biggl\{- \dfrac{a_0(2+\beta+3\lambda)^2}{144  (\beta+\lambda)}\dfrac{\dot{M}^2}{M(t)^2}\biggl[4(r_0^3-r^3)+3r^3\ln\left(\dfrac{r}{r_0}\right)\biggr]\notag\\
& - \dfrac{\dot{M}^2}{36r(\beta+\lambda)M(t)}\sqrt{\dfrac{a_0 G_N}{M(t)}}\biggl[(2+\beta+3\lambda) \times\notag\\&\times\biggl(4r^3 +14r_0^3 - 3r^3\ln\left(\dfrac{r}{r_0}\right)\biggr)+ 36(\beta-1)r_0^3\biggr]\biggr\}\times\notag\\&\times \left[1+{\cal O}(\alpha_1,\alpha_2)\right]+{\cal O}(\beta+\lambda)^0+{\cal O}_{\rm finite}(\dot{M}, \Lambda_{\rm obs})+O(4)\,.
\end{align}
Clearly, if $\dot{M}\neq0$ and $\beta+\lambda\to 0$, the 1PN terms in this expression will dominate over the Newtonian ones, spoiling the agreement
with galaxy rotation curves and with the Tully-Fisher and Faber-Jackson relations.
In the next section we will determine exactly for what values of $\beta+\lambda$ this happens.

\begin{figure*}[t]
\begin{minipage}[b!]{0.4\textwidth}

\begin{tikzpicture}
    \node[anchor=south west,inner sep=0] (image) at (0,0) {\includegraphics[width=\textwidth]{figureHL_new}};
    \begin{scope}[x={(image.south east)},y={(image.north west)}]
        \draw[black,thick, radius=0.18cm] (0.61,0.61) circle ;
        \draw[black,dashed] (0.60,0.635) -- (1.198,0.95) ;
        \draw[white,thick] (0.95,0.963) -- (0.95,0.80) ;
        \draw[black,dashed] (0.59,0.598) -- (1.192,0.151) ;
        
        \draw[black,thick, radius=0.25cm] (0.19,0.58) circle ;
        \draw[black,dashed] (0.18,0.61) -- (0.59,0.96) ;
        \draw[black,dashed] (0.18,0.55) -- (0.59,0.585) ;
        \draw[white,ultra thick] (0.585,0.963) -- (0.95,0.963) ;
        
    \end{scope}
\end{tikzpicture}
\end{minipage}
\begin{minipage}[bh!]{0.4\textwidth}

\begin{tikzpicture} 
    \node[anchor=south west,inner sep=0] (image) at (0,0) {\includegraphics[width=\textwidth]{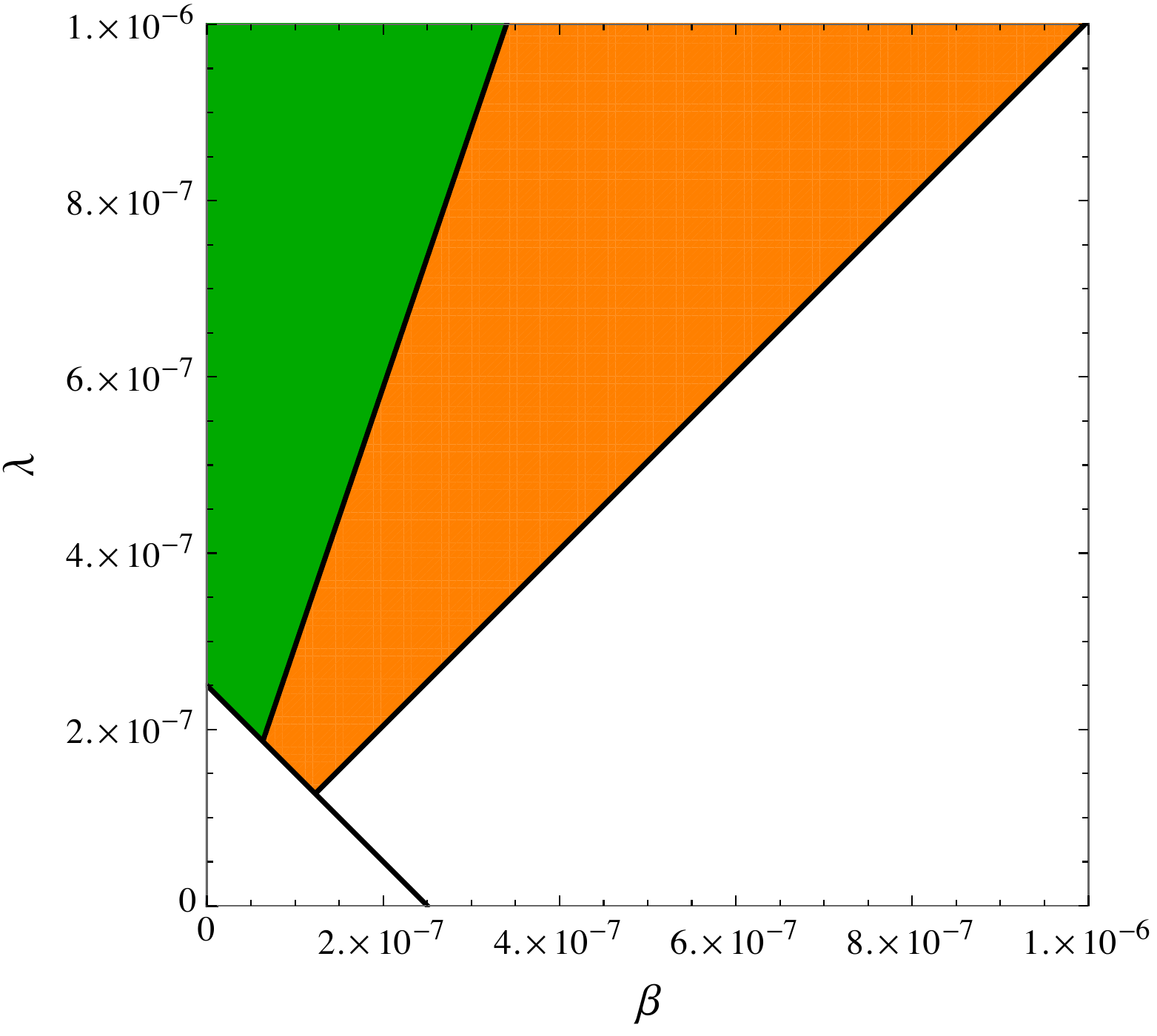}};
    \begin{scope}[x={(image.south east)},y={(image.north west)}]
       
    \end{scope}
\end{tikzpicture}

\end{minipage}
\caption{The bound on $\beta+\lambda$ coming from the requirement that the theory reproduce the rotation curves of galaxies, combined with
constraints from the high-acceleration regime (solar-system tests, absence of \v Cerenkov vacuum radiation, BBN, pulsars, classical and quantum stability).}
\label{fig:vincoli_io}
\end{figure*}

\section{Constraints from the low-acceleration regime}
\label{sec:A new constraint}

In order to determine, at least approximately, the range of the combination $\beta+\lambda$ for which
the PN expansion remains perturbative and the agreement with observations of galactic rotation curves (as well as
with the observed Tully-Fisher relation in disk galaxies and the Faber-Jackson relation in ellipticals and clusters)
is not ruined, let us consider systems (galaxies or clusters) with baryonic masses in the range $M=10^{10} - 10^{14} M_{\odot}$.
Note that for these masses, the radius $r_0$ marking the onset of the MOND effects lies well outside the 
half-light radius given by Eq.~\eqref{eq:fit}, so our calculations (which assume that $r_0$ is larger than the size
of the luminous component of the system) do hold, at least to first approximation.
One crucial ingredient to calculate the impact of the 1PN terms on the rotation curves is, as can be seen
from Eq.~\eqref{eq:velocita_1PN}, the accretion rate of IGM gas onto the galaxy. A very rough estimate for this
quantity is $\dot M\sim M/t_{H}$, where
$t_H\approx 1.4\times 10^{10}$ yr is the Hubble time.

A useful measure of the impact of the 1PN terms on the rotation curves is given by the fractional deviation
\begin{equation}
\epsilon(M,r,\beta+\lambda)=\left\vert v_{\varphi, \mbox{\tiny{1PN}}}^2/v_{\varphi, \mbox{\tiny{N}}}^2 -1 \right\vert\,.
\end{equation}
Clearly, this quantity is a function of $\beta+\lambda$, but also of the galaxy's mass $M$ and the orbital radius $r$. Since 
what is of interest to us is the range of $\beta+\lambda$ for which $\epsilon(M,r,\beta+\lambda)$ is not ``too large'',
we can marginalize over $M\in [10^{10}:10^{14}] M_\odot$, and over $r$. For the latter, we marginalize over a range spanning from
$r=r_0(M)$ (the distance from the center at which MOND effects become dominant) up to the virial radius $r=r_{\rm vir}(M)$ (at $z=0$)
of the $\Lambda$CDM halo corresponding to the galaxy under consideration. This choice is justified because rotation curves are
measured well beyond the galaxy's luminous part, deep into what in the $\Lambda$CDM model is the dark-matter halo region.
In order to estimate $r_{\rm vir}$ (at $z=0$), we use~\cite{Lacey:1993iv,Bryan:1997dn,Klypin:2010qw}
\begin{equation}\label{eq:r_virial}
r_{\rm vir} = \left(\dfrac{M}{f_{\rm b} \times 5.5 \times 10^{13}M_{\odot}}\right)^{1/3}\; \text{Mpc},
\end{equation}
where $f_b\approx 0.17$ is the baryon fraction in the $\Lambda$CDM model.

In order to identify the range of $\beta+\lambda$ for which 1PN terms ``spoil'' the agreement with observations, we
then consider the marginalized fractional deviation
\begin{equation}
\bar{\epsilon}(\beta+\lambda)=\max{\left[\epsilon(M,r,\beta+\lambda)\right]}\Big\vert_{M\in [10^{10}-10^{14} M_\odot]}^{r\in[r_0(M), r_{\rm vir}(M)]}
\,,
\end{equation}
and when 
this quantity exceeds a certain threshold, we conclude that
the 1PN terms jeopardize the agreement between the theory and observations.
Assuming a 20\% threshold (i.e. $\bar{\epsilon}=0.2$), we find that in order for the theory to reproduce galaxy rotation curves, one
must have 
\begin{equation}\label{eq:bound}
|\beta+\lambda|\gtrsim 2.5\times 10^{-7}\,.
\end{equation}
This bound is very conservative, e.g. when considering a 30\% threshold, and marginalizing
only over $M\in [10^{11}:10^{13}] M_\odot$ and $r\in[2 r_0, 0.5 r_{\rm vir}(M)]$, a larger region
of the $(\beta,\lambda)$ plane would remain viable, namely $|\beta+\lambda|\gtrsim 2.5\times 10^{-9}$.
Nevertheless, even with the conservative bound given by Eq.~\eqref{eq:bound}, a significant region of
 the $(\beta,\lambda)$ plane remains viable when one combines that bound with the constraints from the high-acceleration regime
(c.f. discussion in section~\ref{sec:Constraints}). This viable region of the parameter plane is represented in Fig.~\ref{fig:vincoli_io}.

\section{Discussion: open questions and problems}
\label{sec:discussion}
In this paper, we have introduced a theory that can reproduce the MOND phenomenology (and in particular the rotation curves of galaxies)
at low accelerations (i.e. low energies), and which reduces to khronometric theory/Ho\v rava gravity at intermediate/high accelerations (i.e. intermediate/high energies), thus
satisfying experimental requirements such as solar-system tests~\cite{Foster:2005dk,Blas:2010hb,Blas:2011zd}, binary- and isolated-pulsar constraints~\cite{Yagi:2013ava,Yagi:2013qpa}, BBN~\cite{Blas:2012vn,Carroll:2004ai,Yagi:2013ava,Yagi:2013qpa},
the existence of well-behaved black-hole solutions forming from gravitational collapse~\cite{Garfinkle:2007bk,Eling:2006ec,Barausse:2011pu,Blas:2011ni,Barausse:2012ny,Barausse:2012qh,Barausse:2013nwa}, and the absence of
gravitational \v Cerenkov radiation~\cite{Elliott:2005va}, as well as theoretical requirements such as classical and quantum stability~\cite{Blas:2010hb,Blas:2011zd,Garfinkle:2011iw}.

This transition from a MOND-like theory to khronometric theory/Ho\v rava gravity is achieved by making one of the coupling constants of
the theory effectively 
energy-dependent. This was first proposed in Ref.~\cite{Blanchet:2011wv}, but here we generalize that idea
by showing that the theory's 1PN dynamics becomes strongly coupled at low accelerations, unless the other (two) coupling constants of
khronometric theory also have non-zero values. In other words, we show that in order to make the 1PN dynamics perturbative at low energies, the theory 
cannot reduce exactly to GR at intermediate/high energies (as was conjectured by Ref.~\cite{Blanchet:2011wv}), but rather to khronometric theory/Ho\v rava gravity. 
Of course, one clear shortcoming of our approach is that it is purely phenomenological at this stage. Indeed, we \textit{assume} that the running of the coupling constants
is exactly the one that we need to reproduce data/observations. It  remains to be seen if this running is actually the one predicted by the renormalization-group flow,
but as far as we are aware no studies in this direction are available yet.

Another open question about our approach (and about Lorentz-violating gravity in general) is the nature of the mechanism preventing the violations of
Lorentz symmetry from percolating into the matter sector, where they are strongly constrained by cosmic-ray/particle-physics experiments. 
In particular, the higher-order operators that are crucial for the
power-counting renormalizability of Ho\v rava gravity must become important at energies $\lesssim 10^{16}$ GeV to ensure that the theory remains perturbative in the
UV. This scale is comparable with the energy at which Lorentz violations can be probed in the matter sector, thanks to the synchrotron emission from the Crab Nebula~\cite{Liberati:2012jf}.
The percolation of Lorentz violations into the matter sector can of course be suppressed at tree level (by assuming that matter does not couple directly to the Lorentz-violating field),
but it naturally reappears due to radiative corrections. To ensure the viability of the theory, a more efficient suppression mechanism must therefore
be present. Proposals include fine-tuning, ``gravitational confinement''~\cite{Pospelov:2010mp}, ``custodial symmetries'' (e.g. softly broken supersymmetry~\cite{GrootNibbelink:2004za,Pujolas:2011sk}), or dynamical emergence of Lorentz symmetry at low energies in the matter sector, e.g. due to renormalization group flows~\cite{Chadha:1982qq,Bednik:2013nxa}. 

At a more phenomenological level, a pertinent question is whether the theory that we propose can explain all cosmological data (besides galaxy
rotation curves) with no Dark Matter \textit{at all}.
This seems unlikely because MOND itself, as mentioned earlier, requires \textit{some} amount of Dark Matter in the center of galaxy clusters 
-- with mass roughly twice that of observed baryons~\cite{Famaey:2011kh}. 
As mentioned, however, this ``missing mass'' problem is much less serious than in the $\Lambda$CDM model, since
one can postulate that this Dark Matter is given by a (small) fraction of the ``dark missing baryons'' predicted by BBN and not yet observed. In particular,
these dark baryons may be in the form of molecular hydrogen~\cite{2009A&A...496..659T}. Another possibility is that the missing mass in
clusters may be given by neutrinos~\cite{Famaey:2011kh}. (Note that the bounds on the neutrino masses and families from the CMB do not
hold in MOND, at least rigorously, as they assume the $\Lambda$CDM model to start with.) Also, we recall that without some amount of Dark Matter (in baryons
or other components), MOND might have a hard time reproducing observations of the ``Bullet Cluster''~\cite{Markevitch:2003at}, although the interpretation of
the data may be more subtle than initially thought, since a similar system -- the ``Train wreck Cluster''~\cite{Mahdavi:2007yp} -- shows a different behavior.

On scales even larger than those of galaxy clusters (i.e. those relevant for type-Ia supernovae, CMB and large-scale galaxy surveys), the 
full relativistic theory has to be used, in order to account for both the background expansion and perturbations about it. 
For a Robertson-Walker (i.e. homogeneous and isotropic) background, and assuming that the khronon field is aligned with the cosmic time (i.e. that
 hypersurfaces of constant khronon are also ones of constant cosmic time), our theory predicts
the same Friedmann-Lema\^{i}tre-Robertson-Walker equations as in GR, with the only differences being
that \textit{(i)} no Dark Matter is present (except possibly the small amount, 
e.g. in ``dark missing baryons'', needed to explain galaxy-cluster data); and \textit{(ii)} the gravitational constant differs
from the value $G_N$ measured in the solar-system, and is given by $G_C=G_N (1-\alpha/2)/(1+\beta/2+3\lambda/2)$. Given the constraints
on $\alpha$, $\beta$ and $\lambda$ discussed in this paper, $G_C\approx G_N$ to within a few percent. 
This probably makes it difficult to reproduce both type-Ia and CMB data. Indeed type-Ia supernova observations
are only sensitive to the background expansion history, and (to first approximation) constrain a linear combination of 
the density parameters of matter ($\Omega_m$) and cosmological constant ($\Omega_\Lambda$) at $z=0$. 
As for the CMB, a detailed study of perturbations over the cosmological background is needed to
predict the details of its angular spectrum, but the position of its first peak only depends, to first approximation,
on the sound speed of the photon-baryon fluid, and on the angular distance to the baryon-photon decoupling. Both these quantities are the same in
our theory as in the $\Lambda$CDM model. Therefore, because the position of the first CMB peak within the $\Lambda$CDM model constrains $\Omega_m+\Omega_\Lambda\approx 1$
(a constraint almost orthogonal to that coming from type-Ia supernovae), it is clear that our model may have a hard time reproducing both CMB and type-Ia supernova data,
unless we allow as much Dark Matter as in the $\Lambda$CDM model.
A more detailed analysis, however, is needed to confirm this, and will be performed in future work. Indeed, one
may be able to reproduce the data without Dark Matter, but by relaxing the assumption that the khronon must be aligned with the cosmic time.

Another possibility comes from the observation that an effective
 Dark-Matter component on large cosmological scales 
naturally arises in theories similar to ours, namely in
Ho\v rava gravity with the projectability condition. That is a theory with (infrared) action given by Eq.~\eqref{eq:khrono-action-covariant}, but
with $\alpha=0$ and the extra condition (``projectability'') that the lapse $N$ is only a function of time (i.e. $a^\mu=0$) at the level
of the action.
More specifically, Ref.~\cite{Mukohyama:2009mz} (c.f. also Ref.~\cite{Blas:2009yd}) showed that such an effective Dark-Matter component appears
in projectable Ho\v rava gravity if 
deviations from homogeneity are present on large (even super-horizon) scales.
It is also well known that the solutions to projectable Ho\v rava gravity can be obtained from solutions to khronometric theory [action given by Eq.~\eqref{eq:khrono-action-covariant}] in the
limit $\alpha\to\infty$~\cite{Blas:2009qj,Jacobson:2013xta}, or equivalently from solutions to our theory [action given by Eq.~\eqref{action-full-our-theory-covariant}] 
for $\chi\to\infty$.\footnote{We thank Niayesh Afshordi
for suggesting this point.}
This can be shown by following the argument of Ref.~\cite{Jacobson:2013xta}.

 Let us then assume that at the scales relevant for galaxies and clusters we still have $f(a)\approx -2\Lambda_0+2a^2-4 a^3/(3 a_0)$ 
(so that the results of this paper remain valid), but on larger cosmological scales (i.e. even smaller accelerations $a\to0$) 
$f(a) \approx -2\Lambda_0+ {\cal}O(a)$, so that $\chi\propto 1/a$ diverges as $a\to 0$. With this Ans\"atz, the terms depending on $a$ in 
the field equations \eqref{eq:hamiltonian_constraint}--\eqref{eq:evolution_equation_trace} all vanish when $a\to0$, with the exception of the terms giving
the cosmological constant and the term
$D_i (\chi a^i)$ in the modified Hamiltonian constraint \eqref{eq:hamiltonian_constraint}.  
To find the effective Friedmann-Lema\^{i}tre-Robertson-Walker equations in an inhomogeneous universe, one can
take a spatial average of the field equations. In the case of Eq.~\eqref{eq:hamiltonian_constraint}, the average
of the term $D_i (\chi a^i)$ produces a boundary term $C$, which may not be zero if the universe is inhomogeneous
on large (even superhorizon) scales. Indeed, this boundary term might be interpreted as an effective Dark-Matter component, 
because it has the right scaling with  the expansion parameter $A(t)$, i.e.
the spatial average of the modified Hamiltonian constraint \eqref{eq:hamiltonian_constraint} yields
an effective Friedmann-Lema\^{i}tre-Robertson-Walker equation $\dot A^2+ k c^2=8 \pi G A^2 (\rho+\rho_{\rm dm})/3$, with $\rho_{\rm dm}\equiv C/A^3$.
Note that this effective Dark-Matter component might also improve the agreement of the theory with galaxy-cluster
observations, which as mentioned above show some tension with MOND.

Finally, another possibility would be to
replace the term $\theta^2$ in the action~\eqref{action-full-our-theory-covariant} with a function of $\theta^2$. Since $\theta$ is essentially given
by the Hubble rate for a cosmological background, this change may provide enough freedom to fit the background's expansion history, possibly
even providing an effective ``Dark Energy'' component.
(Note that this is similar to the ``generalized'' Einstein-\AE ther theories introduced in Refs.~\cite{Zlosnik:2006sb,Zlosnik:2006zu} or the ``K-essence''
of Ref.~\cite{2010PhLB..684...85G}.) 
Clearly, such a modification of the action~\eqref{action-full-our-theory-covariant}
may affect the analysis of the 1PN dynamics that we performed in this paper, but the formalism that we developed here
is readily extensible to that case.

Of course, all of these possibilities require further detailed exploration
before one can make any definitive claims about them. We will study them, 
both at the level of the cosmological background and perturbations about it, in subsequent publications.

\begin{acknowledgments} During the course of this work we have benefited from inspiring and insightful conversations
and discussions with several colleagues, including 
Niayesh Afshordi, Luc Blanchet, Diego Blas, Monica Colpi, Gilles Esposito-Farese, Ted Jacobson, Luis Lehner, and Shinji Mukohyama.
We also thank Diego Blas, Gilles Esposito-Farese, Ted Jacobson and Luis Lehner for going through a draft of this manuscript and providing useful feedback.
 E.B. acknowledges support 
from the European Union's Seventh Framework Program (FP7/PEOPLE-2011-CIG) through the Marie Curie Career Integration Grant GALFORMBHS PCIG11-GA-2012-321608.
Both M.B. and E.B. acknowledge hospitality from the Lorentz Center (Leiden, NL), where part of this work was carried out. 
\end{acknowledgments}

\section*{Appendix: the PPN parameters in the high-acceleration regime}

\renewcommand{\theequation}{A\arabic{equation}}
\setcounter{equation}{0}

In this Appendix, we show how to solve the 1PN dynamics in the high-acceleration regime, where our theory reduces to
khronometric theory/Ho\v rava gravity. In particular, we confirm, as already shown in Ref.~\cite{Blas:2010hb,Blas:2011zd}, that all the PPN parameters
of khronometric theory are the same as in GR, with the exception of the preferred-frame parameters $\alpha_1$ and $\alpha_2$.

At high accelerations (where $\chi=\alpha$), Eq.~\eqref{eq:modified poisson equation with chi} yields the usual expression for the Newtonian potential, 
\begin{equation}
\phi_N = -G_N\int d^3 x' \dfrac{\rho(\vec{x}\ ', t)}{|\vec{x} - \vec{x}\ '|}\,,
\end{equation}
where we recall that the locally measured gravitational constant $G_N$ is related to the ``bare'' one appearing in the action by Eq.~\eqref{GN}.
The equations characterizing the 1PN dynamics are Eqs.~\eqref{eq:modified_poisson_1PN} and \eqref{eq:momentum constraint 1PN}, which in the high-acceleration regime become
\begin{align}
\label{eq:poisson_1PN high acceleration}
&\nabla^2 \biggl(\phi_N + \dfrac{\phi_{\mbox{\tiny{(2)}}}}{c^2}\biggr) = 4\pi G_N \rho\notag\\
&\qquad + \dfrac{1}{c^2}\Bigl[8\pi G_N \rho v^2 + 12\pi G_N p + 2\vec{\nabla}\phi_N\cdot\vec{\nabla} \phi_N \notag\\
&\qquad- \dfrac{2+\beta+3\lambda}{2-\alpha}\bigl(\partial_t\nabla^2 \omega + 3\partial_t^2\phi_N\bigr)\Bigr].\\\label{eq:momentum constraint 1PN high acceleration}
&\nabla^2 w_i + 2\Bigl(\dfrac{\beta+\lambda}{\beta -1}\Bigr)\partial_i \nabla^2 \omega\notag\\
&\qquad = \dfrac{16\pi G \rho v_i }{1-\beta} -2\Bigl(\dfrac{2+\beta+ 3\lambda}{\beta -1}\Bigr)\partial_i \partial_t\phi_N\,.
\end{align}
Before solving them, let us first define the PN potentials~\cite{Will:1993ns}:
\begin{gather}
\mathbb{X}(\vec{x}, t) =  G_N \int d^3 x' \rho(\vec{x} ', t) \ |\vec{x} - \vec{x}'|, \\ V_{i}=G_N\int d^{3}x' \ \frac{\rho(\vec{x} ', t) v'_{i}}{|\vec{x}-\vec{x} '|},
\\
 W_{i}=G_N\int d^{3}x' \ \frac{\rho(\vec{x} ', t) \ \vec{v} ' \cdot (\vec{x}-\vec{x}\ ') \ (x-x')_{i}}{|\vec{x}-\vec{x} '|}, \\ 
\Phi_{1}=G_N\int d^{3}x' \frac{\rho(\vec{x} ', t) v'^{2}}{|\vec{x}-\vec{x}'|}, \\\Phi_{2}=-G_N\int d^{3}x' \frac{\rho(\vec{x} ', t) \phi_N(\vec{x} ', t)}{|\vec{x}-\vec{x} '|}, \\ \Phi_{4}=G_N\int d^{3}x' \frac{p(\vec{x} ', t)}{|\vec{x}-\vec{x} '|}.
\end{gather}
and recall the following relations among them~\cite{Will:1993ns}:
\begin{gather}\label{eq:relazioni1}
 \nabla^2 \mathbb{X} = -2\phi_N, \\ \nabla^2 V_i = -4\pi G_N \rho v_i, \\ \nabla^2\Phi_{1} = -4\pi G_N \rho v^2, \\ \nabla^2\Phi_{2} = 4\pi G_N \rho \phi_N,\\
\nabla^2\Phi_{4} = -4\pi G_N p, \\ \partial_i V^i = \partial_t \phi_N, \\ \partial_i V^i = -\partial_i W^i, \\  \partial_t\partial_i \mathbb{X} = W_i - V_i\,.\label{eq:relazioni_last}
\end{gather}
Equation~\eqref{eq:momentum constraint 1PN high acceleration} can then be written as
\begin{align}
&\nabla^2 w_i + 2\Bigl(\dfrac{\beta+\lambda}{\beta -1}\Bigr)\partial_i \nabla^2 \omega=\notag\\
&  -\dfrac{2 (2-\alpha) \nabla^2 V_i }{1-\beta} +\Bigl(\dfrac{2+\beta+ 3\lambda}{\beta -1}\Bigr)\partial_i \partial_t\nabla^2\mathbb{X}\,.\label{fd_eq}
\end{align}
Taking the divergence of this equation and using the relations above between the PN potentials, we then obtain the solution for $\omega$, i.e.
\begin{equation}
\omega = \dfrac{\alpha + \beta +3\lambda}{2(\beta+\lambda)}\partial_t \mathbb{X}\,,
\end{equation}
which, when replaced back in Eq.~\eqref{fd_eq}, allows the computing of $w_i$. The solution for $g_{0i}$ then reads 
\begin{align}
g_{0i} &= \dfrac{w_i}{c^3} + \dfrac{\partial_i \omega}{c^3} +O(5)\notag\\&= \dfrac{\beta^2 + \lambda + 3\beta(1+\lambda)-\alpha(1+\beta+2\lambda)}{2(\beta-1)(\beta+\lambda)}\dfrac{W_i}{c^3}\notag\\
& + \dfrac{\alpha+5\beta-3\alpha\beta-\beta^2+\lambda(7-2\alpha-3\beta)}{2(\beta-1)(\beta+\lambda)}\dfrac{V_i}{c^3} +O(5).
\end{align}
By using the solution for $\omega$ and the relations \eqref{eq:relazioni1}-\eqref{eq:relazioni_last}, one can then solve  
Eq. \eqref{eq:poisson_1PN high acceleration} for $\phi_{\mbox{\tiny{(2)}}}$, obtaining
\begin{multline}
\phi_{\mbox{\tiny{(2)}}} = \phi_N^2 - 2\Phi_1 - 2\Phi_2 - 3\Phi_4 \\+ \dfrac{(\alpha-2\beta)(2+\beta+3\lambda)}{2(\alpha-2)(\beta+\lambda)}\partial^2_t \mathbb{X}\,,
\end{multline}
which yields the complete solution for $g_{00}$ at 1PN order:
\begin{align}
g_{00} &= -1 -2\dfrac{\phi_N}{c^2} -2\dfrac{\phi_N^2}{c^4} + 4\dfrac{\Phi_1}{c^4} + 4\dfrac{\Phi_2}{c^4} + 6\dfrac{\Phi_4}{c^4}\notag\\
& - \dfrac{(\alpha-2\beta)(2+\beta+3\lambda)}{(\alpha-2)(\beta+\lambda)}\dfrac{\partial^2_t \mathbb{X}}{c^4} +O(6).
\end{align}
Finally, by performing a gauge transformation $t\to t+\delta t$ (with $\delta t\propto \partial_t\mathbb{X}$), 
we can write the 1PN metric in the standard PN gauge, i.e.
\begin{align}
&g_{00} = -1 -2\dfrac{\phi_N}{c^2} -2\dfrac{\phi_N^2}{c^4} + 4\dfrac{\Phi_1}{c^4} + 4\dfrac{\Phi_2}{c^4} + 6\dfrac{\Phi_4}{c^4}+O(6)\\
& g_{0i} =-\dfrac{1}{2}\Bigl(7 +\alpha_1-\alpha_2\Bigr)\dfrac{V_i}{c^3} - \dfrac{1}{2}\Bigl(1 +\alpha_2\Bigr)\dfrac{W_i}{c^3}+O(5)\\
& g_{ij} = \Bigl(1-2\dfrac{\phi_N}{c^2}\Bigr)\delta_{ij}+O(4)
\end{align}
where the preferred frame parameters are given, as in Refs.~\cite{Blas:2010hb,Blas:2011zd}, by
\begin{align}
\alpha_1&=\dfrac{4(\alpha - 2\beta)}{\beta-1},\\
\alpha_2&=\dfrac{(\alpha-2\beta)[-\beta(3+\beta+3\lambda) - \lambda +\alpha(1+\beta+ 2\lambda)]}{(\alpha-2)(\beta-1)(\beta+\lambda)}\,.
\end{align}

\bibliography{Bibliografia} 

\end{document}